\documentclass{aa}  

\usepackage{graphicx}
\usepackage{txfonts}
\usepackage{textcomp}
\usepackage{float}
\begin{document}

\title{Young, active radio stars in the AB\,Doradus moving group}

   %\subtitle{I. Overviewing the $\kappa$-mechanism}

\author{R. Azulay
\inst{1,2}\fnmsep\thanks{Guest student of the International Max Planck Research School for Astronomy and Astrophysics at the Universities of Bonn and Cologne},
J. C. Guirado\inst{3,1},
J. M. Marcaide\inst{1},
I. Mart\'i-Vidal\inst{4},
E. Ros\inst{2,1,3},
E. Tognelli\inst{5,6},
F. Hormuth\inst{7},
\and
J.~L.~Ortiz\inst{8}
}

\institute{Departament d'Astronomia i Astrof\'{\i}sica, Universitat de
Val\`encia, C. Dr. Moliner 50, E-46100 Burjassot, Val\`encia, Spain\\
\email{Rebecca.Azulay@uv.es}
\and
Max-Planck-Institut f\"ur Radioastronomie, Auf dem
H\"ugel 69, D-53121 Bonn, Germany
\and
Observatori Astron\`omic, Universitat de Val\`encia, Parc
Cient\'{\i}fic, C. Catedr\'atico Jos\'e Beltr\'an 2, E-46980
Paterna, Val\`encia, Spain
\and
Onsala Space Observatory, Chalmers University of Technology, SE-43992 Onsala, Sweden
\and
Department of Physics `E.Fermi', University of Pisa, Largo Bruno Pontecorvo 3, I-56127 Pisa, Italy
\and
INFN, Section of Pisa, Largo Bruno Pontecorvo 3, I-56127 Pisa, Italy
\and
Max-Planck-Institut f\"ur Astronomie, Koenigstuhl 17, D-69117 Heidelberg, Germany
\and
Instituto de Astrof\'isica de Andaluc\'ia (IAA-CSIC), Apt 3004, E-1808, Granada, Spain
}

   \date{Draft version: \today}

% \abstract{}{}{}{}{} 
% 5 {} token are mandatory
 
  \abstract
%%%  % context heading (optional)
%%%  % {} leave it empty if necessary  
{Precise determination of stellar masses is necessary to test the validity of pre-main-sequence (PMS) stellar evolutionary models, whose predictions are in disagreement with measurements for masses below 1.2\,M$_{\odot}$. To improve such a test, and based on our previous studies, we selected the AB\,Doradus moving group (AB\,Dor-MG) as the best-suited association on which to apply radio-based high-precision astrometric techniques to study binary systems.}
%%%  % aims heading (mandatory)
{We seek to determine precise estimates of the masses of a set of stars belonging to the AB\,Dor-MG using radio and infrared observations.}
%%%  % methods heading (mandatory)
{We observed in phase-reference mode with the Very Large Array (VLA) at 5\,GHz and with the European VLBI Network (EVN) at 8.4\,GHz the stars HD\,160934, EK\,Dra, PW\,And, and LO\,Peg. We also observed some of these stars with the near-infrared CCD AstraLux camera at the Calar Alto observatory to complement the radio observations.} 
%%%  % results heading (mandatory)
{We determine model-independent dynamical masses of both components of the star HD\,160934, A and c, which are 0.70$\pm$0.07\,M$_{\odot}$ and 0.45$\pm$0.04\,M$_{\odot}$, respectively. We revised the orbital parameters of EK\,Dra and we determine a sum of the masses of the system of 1.38$\pm$0.08\,M$_{\odot}$. We also explored the binarity of the stars LO\,Peg and PW\,And.} 
%%%  % conclusions heading (optional), leave it empty if necessary 
{We found observational evidence that PMS evolutionary models underpredict the mass of PMS stars by 10\%$-$40\%, as previously reported by other authors. We also inferred that the origin of the radio emission must be similar in all observed stars, that is, extreme magnetic activity of the stellar corona that triggers gyrosynchrotron emission from non-thermal, accelerated electrons.}

   \keywords{stars: binaries -- stars: pre-main-sequence -- radio emission -- astrometry} 
   \titlerunning{Young, active radio stars in the AB\,Doradus moving group}
    \authorrunning{Azulay et al.}
   \maketitle
%
%________________________________________________________________

\section{Introduction}

%Stellar evolution models allow us to know and to understand the different phases that the stars cross throughout its existence. As a general rule, the predictions that these models provide fit correctly to the observations and, therefore, are a truly reliable source of scientific information. This is particularly useful to estimate fundamental parameters of the stars, as the mass and radius, from theoretical luminosity-based relationships (e.g., Baraffe et al. 1998; Chabrier et al. 2000). The calibration of these stellar evolution models is important, but it is crucial in the case of young, low-mass objects, since these models are deeply used to determine the masses of planets and brown dwarfs. However, in the particular case of pre-main-sequence (PMS) stars with masses below 1.2\,M$_{\odot}$, the efficiency of the models is questionable since the contrast with the observational data reveals some discrepancies (Hillenbrand \& White 2004). Therefore, the calibration of the evolutionary models of low-mass PMS stars can be considered an important task, but certainly is not easy, since it requires precise and independent measurements of luminosities and masses to be compared with the theoretical predictions. 

Stellar evolution models are an essential tool to infer star fundamental parameters such as radius, mass, and/or age from luminosity/temperature-based relationships (e.g., Baraffe et al. 1998; Chabrier et al. 2000). The reliability of the models has long been tested and validated by the overall good agreement between the predictions of stellar models and measurements. However, only recently, accurate measurements of stellar masses and radii have become accessible, especially in the case of low- and very low-mass stars, thus allowing more stringent tests on stellar models. In the particular case of pre-main-sequence (PMS) stars the models show an increasing difficulty in accurately reproducing some of the characteristics of star with masses below 1.2\,M$_{\odot}$ (e.g., Hillenbrand \& White 2004).

Therefore, the calibration of the evolutionary models of low-mass PMS stars is an important and challenging task, since it requires precise and independent measurements of luminosities and masses to be compared with theoretical predictions. Several authors have highlighted these facts in previous works but, nevertheless, there is not enough observational data that can help to improve the models yet (Hillenbrand \& White 2004; Stassun et al. 2004; Mathieu et al. 200; Gennaro et al. 2012).

The study of binary stars belonging to young moving groups, whose main feature is the common age of their members, is a reasonable approach to increase the number of PMS stars with dynamically determined masses. Several of these moving groups have recently been discovered (Zuckerman \& Song 2004; Torres et al. 2008). Among all of these groups, the AB\,Doradus moving group (AB\,Dor-MG) is the most suitable to carry out the study; because this group is the closest moving group, the estimated age is relatively accurate and contains stars with significant emission at radio wavelengths (Guirado et al. 2006, 2011; Jason et al. 2007; Azulay et al. 2014, 2015). This last feature is essential because it allows us to use radio interferometry techniques to obtain astrometric information. Using these techniques it is possible to achieve angular resolutions in the sub-milliarcsecond (sub-mas) range, which are needed to solve and study in detail the kinematics (proper motion, parallax, and possible orbits) of the stellar systems.

In this context, we have made several contributions to stars belonging to the AB\,Dor-MG, namely, AB\,Dor\,A/C (Guirado et al. 2006; 2011), AB\,Dor\,Ba/Bb (Azulay et al. 2015), and HD\,160934 (Azulay et al. 2014). In the two first cases, a VLBI-driven astrometric study resulted in the precise estimate of the dynamical mass of the individual components, providing relevant results in terms of calibration of the mass-luminosity relationship for young, low-mass objects. Regarding the binary HD\,160934, we reported the discovery of compact radio emission from both components of the system, which opened the possibility to a further astrometric monitoring of its orbital motion.

Given the remarkable scientific output of AB\,Dor\,A/C, AB\,Dor\,Ba/Bb, and HD\,160934, we considered it appropriate to include new similar stars, that is, young binaries that are luminous both in infrared and radio wavelengths. In fact, other stars in the AB\,Dor-MG are fast rotators, showing traces of magnetic activity (as stellar spots) that well could be radio emitters. The previous reasoning was the main motivation to initiate a study of the radio emission of AB\,Dor-MG members beyond the systems already studied.

In this paper we present the results of a VLA/VLBI radio study of PMS stars members of the AB\,Dor-MG, namely, HD\,160934, EK\,Dra, PW\,And, and LO\,Peg. In particular, we focus on the VLBI observations of HD\,160934, from which we were able to monitor astrometrically the relative orbit (of the component HD\,160934\,A respect to the component HD\,160934\,c) and the absolute orbit (reflex motion of the component HD\,160934\,A with respect to a external quasar) and, thereby, to determine dynamical individual masses of both components of the star, which enabled further comparisons with stellar models. We also report on VLBI observations of the other three stars addressed to determine their fundamental parameters (EK\,Dra) or explore its possible binarity (PW\,And and LO\,Peg).

\section{Observations and data reduction}

\begin{table*}
\caption{Journal of observations}             
\label{journal}      
\begin{center}
\resizebox{\hsize}{!}{         
%\begin{tabular}{@{}ccccc@{}ccc@{}}
\begin{tabular}{ccccccc}
\hline\hline 
      \noalign{\smallskip}
%\multicolumn{7}{c}{VLA observations} \\ 
%Source & Date (Epoch) & VLA configuration & UT Range & \multicolumn{2}{c}{Beam size} & P.A. \\ 
%       &     &                  &   &  \multicolumn{2}{c}{[arcsec]} & [$^\circ$]~~\\
Source & Date (Epoch) & VLA configuration & UT Range & Beam size & P.A. \\ 
        &     &                  &   &  [arcsec] & [$^\circ$]~~\\
\noalign{\smallskip}
            \hline
            \noalign{\smallskip}
HD\,160934 & 13 Feb 2009 & B & 17:15\,-\,22:15 & 1.01$\times$0.79 & $-$35.9\\
EK\,Dra & 29 Jan 1993 & AB & 01:25\,-\,12:00 & 0.74$\times$0.33 & $-$72.8\\          
PW\,And & 16 Sep 1993 & CD & 06:00\,-\,16:00 & 10.90$\times$3.28 & 80.5\\ 
LO\,Peg & 5 May 1996 & CD & 12:50\,-\,18:10 & 7.63$\times$3.44 & 82.0\\
\hline
      \noalign{\smallskip}
% & & & & & \\
%\multicolumn{7}{c}{VLBI observations} \\ 
%Source & Date (Epoch) & VLBI array$^\mathrm{a}$ & UT Range & \multicolumn{2}{c}{Beam size} & P.A. \\ 
%       &     &                  &   &  \multicolumn{2}{c}{[mas]} & [$^\circ$]~~\\
Source & Date (Epoch) & VLBI array$^\mathrm{a}$ & UT Range & Beam size & P.A. \\ 
        &     &                  &   &  [mas] & [$^\circ$]~~\\
\noalign{\smallskip}
            \hline
            \noalign{\smallskip}  
HD\,160934 & 30 Oct 2012 & Ef, Wb, Jb, On, Mc, Nt, Tr, Ys, Sv, Zc, Bd, Ur, Sh & 10:30\,-\,20:30 & 2.29$\times$1.59 & $-$28.4\\          
" & 23 May 2013 & " & 21:00\,-\,07:00 & 2.20$\times$1.76 & $-$31.3\\
" & 5 Mar 2014 & " & 02:00\,-\,12:00 & 2.05$\times$1.59 & $-$27.3\\ 
EK\,Dra & 29 Oct 2012 & Ef, Wb, Jb, On, Mc, Nt, Tr, Ys, Sv, Zc, Bd, Ur, Sh & 07:30\,-\,17:30 & 2.19$\times$1.71 & $-$34.1\\
" & 27 May 2013 & " & 17:45\,-\,03:45 & 1.30$\times$1.02 & $-$30.5\\
" & 5 Mar 2014 & " & 16:30\,-\,02:30 & 1.79$\times$0.96 & $-$13.0\\
Pw\,And & 26 Oct 2014 & Ef, Wb, Jb, On, Nt, Tr, Ys, Sv, Zc, Bd, Sh & 16:30\,-\,02:30 & 4.63$\times$2.84 & 0.4\\        
LO\,Peg & 23 Oct 2014 & Ef, Wb, Jb, On, Nt, Tr, Ys, Sv, Zc, Bd, Sh, Hh & 13:00\,-\,23:00 & 1.21$\times$0.89 & $-$84.4\\   
\noalign{\smallskip}
\hline
\end{tabular}
}
\end{center}
\footnotesize{\textbf{Notes.} $^\mathrm{a}$: Ef: Effelsberg, Wb: Westerbork, Jb: Jodrell Bank, On: Onsala, Mc: Medicina, Nt: Noto, Tr: Torun, Ys: Yebes, Sv: Svetloe, Zc: Zelenchukskaya, Bd: Badary, Ur: Urumqi, Sh: Shanghai, Hh: Hartebeesthoek.}
\end{table*}

\subsection{VLA observations}
We analyzed archival VLA data\footnote{Projects AG0377, ADA000, and ABO691 available at the VLA data archive https://archive.nrao.edu/archive/advquery.jsp} of the stars EK\,Dra, PW\,And, and LO\,Peg observed at 8.4\,GHz in AB (EK\,Dra) and CD (PW\,And and LO\,Peg) configurations on 1993 January 29, 1993 September 16, and 1996 May 5, respectively (see Table \ref{journal}). In all cases, the effective bandwidth was 50\,MHz and both right- and left-handed circular polarizations were collected. For EK\,Dra the observation lasted 10.5\,h, the source 3C48 was used as the primary flux calibrator and the source 1435+638 was selected as the phase calibrator. For PW\,And, the observation lasted 10\,h, and the flux and phase calibrators were 0137+331 and 0029+349, respectively. For LO\,Peg, the observation lasted 5.5\,h, the source 0137+331 was used as primary flux calibrator, and the source 2115+295 was selected as phase calibrator. VLA observations of HD\,160934 are described in Azulay et al. (2014) and included in Table \ref{journal} for completeness.

To reduce all three experiments, we used standard routines of the Astronomical Image Processing System (AIPS, 31DEC15 version) program of the National Radio Astronomy Observatory (NRAO), which we summarize in turn. We flagged data, both obvious outliers or data segments selected after careful checking of the observing logs, which constituted a small fraction of the complete data set; we determined the flux density of the primary calibrator, we calculated the flux density of the phase calibrator from the primary flux calibrator, and we used the solutions derived from the calibrators to calibrate the amplitudes and phases of the target through linear interpolation. These calibrated data were imported to the \textit{DIFMAP} software package (Shepherd 1994) to obtain the images of the stars. These resulting images are shown in Figs.~\ref{ekdra_vla}, \ref{pwand_vla}, and \ref{lopeg_vla} and are discussed in the next section.
%We analyzed archival VLA data of the stars LO\,Peg and PW\,And observed at 8.4\,GHz in CD configuration on 1996 May 5 and 1993 September 16, respectively. In both cases, the effective bandwidth was 50\,MHz and left and right circular polarizations were collected. For LO\,Peg, the observation lasted 5.5\,h, the source 0137+331 was used as primary flux calibrator, and the source 2115+295 was selected as phase calibrator. For PW\,And, the observation lasted 10\,h, and the flux and phase calibrators were 0137+331 and 0029+349, respectively.

%To reduce both experiments, we used standard routines of the Astronomical Image Processing System (AIPS) program, of the National Radio Astronomy Observatory (NRAO). We flagged bad data, we determined the flux density of the flux calibrator, we calculated the flux density of the phase calibrator from the primary flux calibrator, and we used the solutions derived from the calibrators to calibrate the gains and phases of the target through linear interpolation. These calibrated data were imported to the DIFMAP software-package (Shepherd 1997) to obtain the images of the stars. These resulting images are shown in Fig. \ref{lopeg_vla}, \ref{pwand-vla}. This are the first images at radio wavelengths of LO\,Peg and PW\,And, were we can reveal a flux density of 0.45\,mJy and 0.34\,mJy, respectively.

\subsection{VLBI observations}
The previous VLA observations certified the presence of radio emission on HD\,160934, EK\,Dra, PW\,And, and LO\,Peg; to study their compact structures, we carried out VLBI observations at 5\,GHz between 2012 and 2014 with the EVN (see Table \ref{journal}). Results of the first VLBI epoch on HD\,160934 were already presented in Azulay et al. (2014) but they have been reanalyzed here in the context of new observations. For each experiment we observed an overall time of 10\,h and both polarizations were recorded with a rate of 1024 Mbps (two polarizations, eight sub-bands per polarization, 16 MHz per sub-band, and two bits per sample). After each observation, the data were correlated with the EVN MkIV data processor at the Joint Institute for VLBI in Europe (JIVE), Dwingeloo, The Netherlands.

As we are studying weak sources, we used the phase-referencing technique to facilitate its detection; for that, we interleave scans of the target sources with ICRF quasars. The quasars selected were J1746$+$6226, J1441$+$6318, J0015$+$3216, and J2125$+$2442 for HD\,160934, EK\,Dra, PW\,And, and LO\,Peg, respectively (separated by 1.50$^\circ$, 1.04$^\circ$, 1.48$^\circ$, and 1.87$^\circ$, respectively). The cycles target-calibrator-target lasted about six minutes in all cases.

\begin{figure*}
\resizebox{\hsize}{!}
{\includegraphics{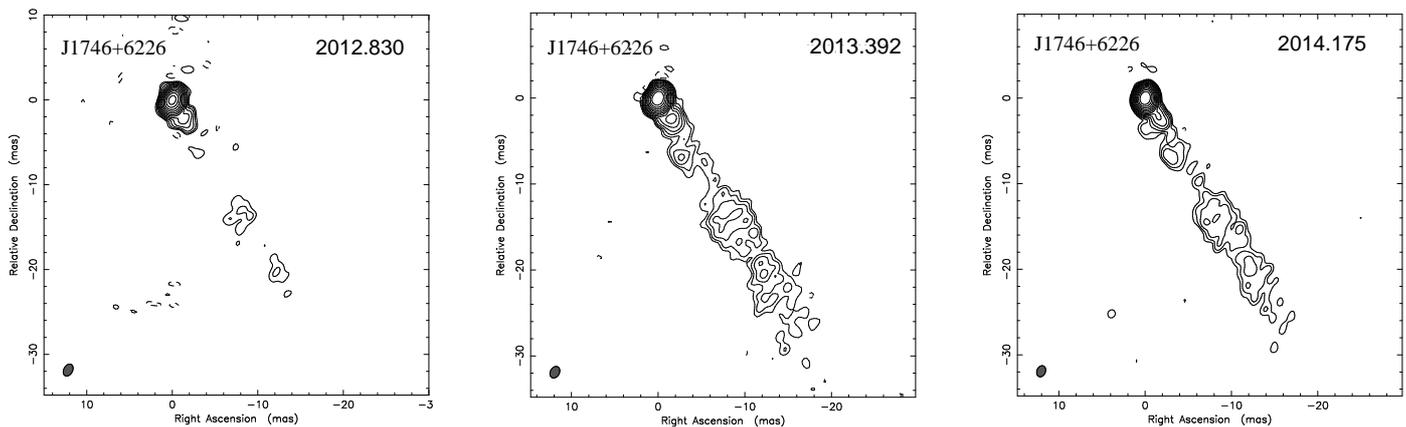}}
\caption[EVN maps of the calibrator J1746$+$6226]{Maps of J1746$+$6226 (calibrator of HD\,160934) in the three epochs. In each map, the lowest contour levels correspond to 4 times the statistical rms noise (0.5, 0.2, and 0.2\,mJy\,beam$^{-1}$) with a scale factor between contiguous contours of $\sqrt{3}$. The peak flux densities in the images are 0.32, 0.32, and 0.34 \,mJy\,beam$^{-1}$, respectively. The different morphology in epoch 2012.830 is due to the larger rms.}
\label{j1746}
\end{figure*}

We reduced each experiment using AIPS in a standard procedure briefly described here. The initial reduction included amplitude calibration using system temperatures and antenna gains, corrections of parallactic angle, and ionosphere. We applied fringe fitting on the calibrator to determine phase offsets and applied the solutions to the target source. Later, we imported the resulting data to \textit{DIFMAP} to obtain uniformly weighted maps of the calibrator (Fig. \ref{j1746}, \ref{j1441}, \ref{j0015}, and \ref{j2125}). We obtained the image through a process of self-calibration iterations of the amplitude and phase with deconvolutions using the \textsc{clean} algorithm, which allowed us to determine both the amplitude scaling corrections and self-calibrated phase for each antenna. Back to AIPS, we applied these corrections to the target source to obtain the phase-referenced naturally weighted images (Fig. \ref{hd}, \ref{ekdra}, and \ref{pwand}). We analyze the details in the next section.

%We reduced each experiment using AIPS in a standard procedure. After making the initial reduction (that includes the amplitude calibration using system temperatures and antenna gains, the corrections of the parallactic angle and the variations due to the ionosphere, and the fringe-fitting on the calibrator to determine phase offsets and applied the solutions to the target source) we imported the calibrated data to DIFMAP to obtain uniformly-weighted maps of the calibrator (Fig. \ref{j1746}, \ref{j1441}, \ref{j2125}, \ref{j0015}). We obtain the image through a process of iteration of the amplitude and phase self-calibrations with deconvolutions using the \textsc{clean} algorithm, that allowed us to determine both the amplitude scaling corrections and self-calibrated phase for each antenna. Back to AIPS, we applied these corrections to the target source to obtain the phase-referenced naturally-weighted image (Fig. \ref{hd}, \ref{ekdra}, \ref{pwand}). %Circular Gaussian least-squares-fitted parameters for the components are listed in Table xx.

\subsection{AstraLux observations}
HD\,160934 and EK\,Dra were also observed with the Lucky Imaging AstraLux camera (Hormuth et al.~2008) at the Calar Alto 2.2\,m telescope. The Lucky Imaging technique permits the reducuction of distortions due to atmosphere by acquiring a large number of short-exposure images and combining the best few percent of high-quality images to obtain a final image that is relatively unaffected by atmospheric turbulence (Hormuth et al.~2007 and reference therein). The observations of HD\,160934 were carried out on 2013 June 24 and 2015 November 19; the observations of EK\,Dra were carried out on 2013 June 24 2013. In all cases, we used two different filters, SDSS i$^\prime$ and SDSS z$^\prime$, and we took 10000 individual frames with exposure times of 30\,ms each. The individual frames were dark and flat corrected before selecting the best 10\% of the accquisitions. The final images were constructed by filtering and resampling the selected frames, and then combining them with the Drizzle algorithm (Fruchter \& Hook 2001). We also observed the stars at the center of the globular cluster M15, whose positions were used for astrometric calibration in the way described in Hormuth et al.~(2007). At each observation, the field of view was 24$^{\prime\prime}\times$24$^{\prime\prime}$ in a 512$\times$512 pixel frame. The resulting images are shown in Fig.~\ref{hd_caha} and \ref{ekdra_caha}. Details are shown in the next section.

\section{Discussion on individual sources}

\subsection{HD\,160934}
HD\,160934 (=HIP\,86346) is a young, very active, binary star (Zuckerman, Song, and Bessel 2004; López-Santiago et al. 2006). The two components, HD\,160934\,A and HD\,160934\,c, were first reported by Galvez et al. (2006). This star has spectral type K7Ve (Schlieder et al. 2012) and it is placed at a distance of $\sim$33\,pc (van Leeuwen 2007).

Several studies of HD\,160934 have been carried out so far with different techniques. Radial velocity measurements were performed by Henry et al. (1995), Zuckerman et al. (2004), Gálvez et al. (2006), Griffin \& Filiz Ak (2010), and Griffin (2013). Relative astrometry through infrared imaging was reported by Hormuth et al. (2007) and Lafrenière et al. (2007). Moreover, relative astrometry from aperture-masking interferometry was provided by Evans et al. (2012). In Azulay et al. (2014), we reported the discovery of the radio emission of both components of the pair. There, we used that discovery to carry out relative astrometry and to determine new orbital elements and mass estimates.

Regarding our new VLBI images, only a single component (assigned to A) is detected in 2013.392 (the upper bound to the radio emission of the component c is 0.01\,mJy). In contrast, in the images corresponding to 2014.175, and as it happened in the observations of 2012.830 reported in Azulay et al. (2014), two point-like features are clearly seen. Those two point-like features can readily be associated with components A and c of the binary HD\,160934. Circular Gaussian least-squares-fitted parameters for the two components are listed in Table \ref{tablegauss_hd}.

\begin{table}
\caption{Circular Gaussian estimates of the components of HD\,160934}            
\label{tablegauss_hd}
\begin{center}
\begin{tabular}{c c c c }     % 7 columns 
\hline\hline       
                       
Epoch & Component & Flux  & Diameter\\    % table heading 
      &    & (mJy) & (mas)\\    % table heading
\hline
2012.830 &  A & 0.16$\pm$0.01 & 2.76$\pm$0.17\\      % inserting body of the table
   & c & 0.06$\pm$0.01 & 1.20$\pm$0.20\\
\hline                   
2013.392 &  A & 0.05$\pm$0.01 & 2.52$\pm$0.46\\      % inserting body of the table
\hline                                   %inserts single line
2014.175 &  A & 0.13$\pm$0.01 & 2.76$\pm$0.20\\      % inserting body of the table
   & c & 0.06$\pm$0.01 & 1.83$\pm$0.30\\
\hline                
\end{tabular}
\end{center}
\end{table}

\begin{figure*}
\begin{center}
\resizebox{0.8\hsize}{!}
{\includegraphics{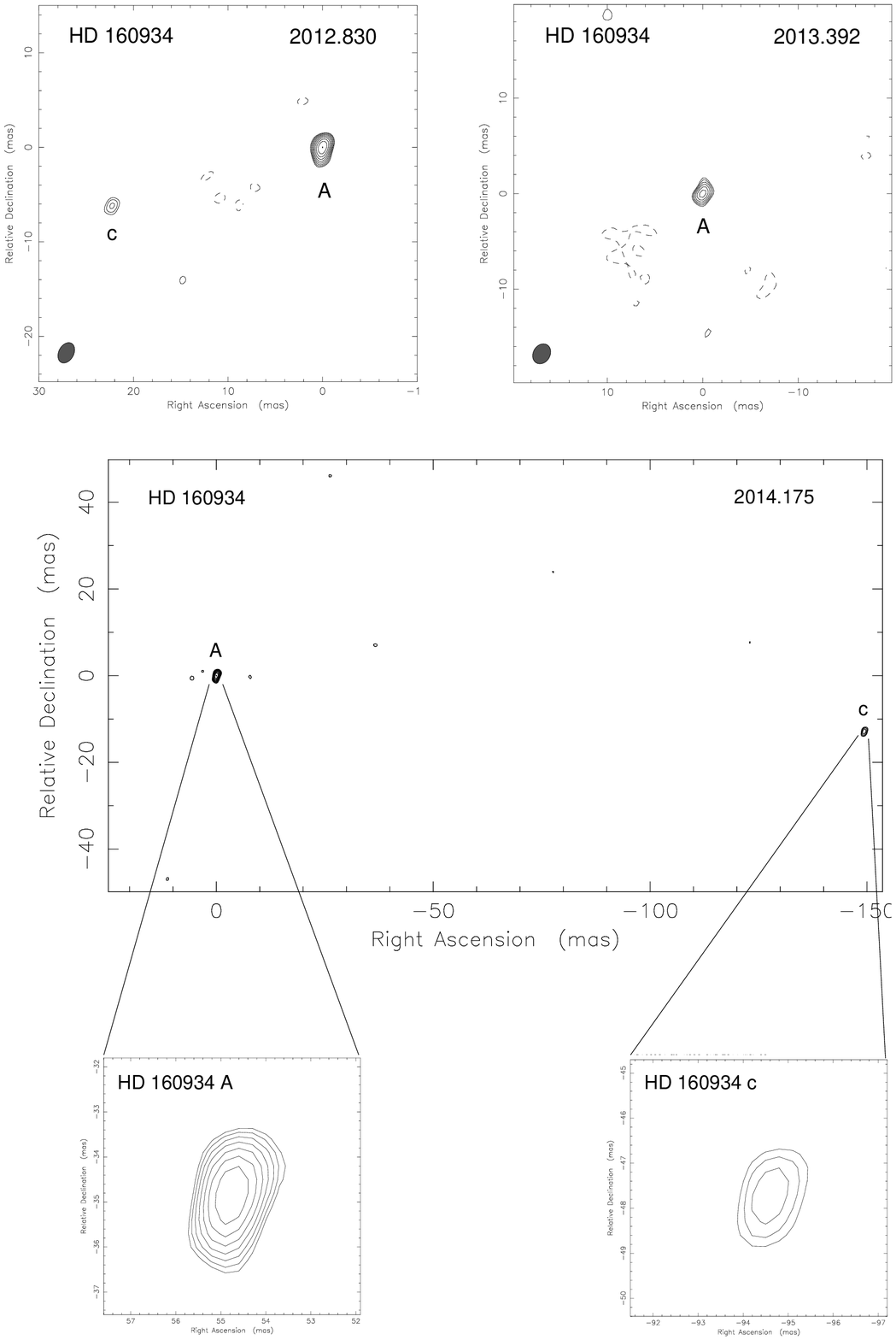}}
\caption[EVN maps of the star HD\,160934]{Clean maps of the binary HD\,160934 at the three EVN epochs. The lowest contour levels corresponds to 3 times the statistical rms noise (0.04, 0.02, and 0.04\,mJy\,beam$^{-1}$) with a scale factor between contiguous contours of $\sqrt{2}$. The peak flux densities in the images are 0.15, 0.04, 0.12\,mJy\,beam$^{-1}$, respectively. For image parameters see Table \ref{journal}. In all the maps we set the origin at the position of the peak of brightness of HD\,160934\,A. Component c is not detected at epoch 2013.392. Orbital motion of c around A is evident from epochs 2012.830 and 2014.175.}
\label{hd}
\end{center}
\end{figure*}

In the AstraLux images, meanwhile, the two components are not distinguished in the 2013 images whereas they are in those of 2015, where c is near the apoastron. We fitted a binary model to find the separation and flux ratio of the latter images; the results are shown in Table \ref{tablehd}, along with previous estimates.

\begin{figure*}
\begin{center}
\resizebox{0.6\hsize}{!}
{\includegraphics{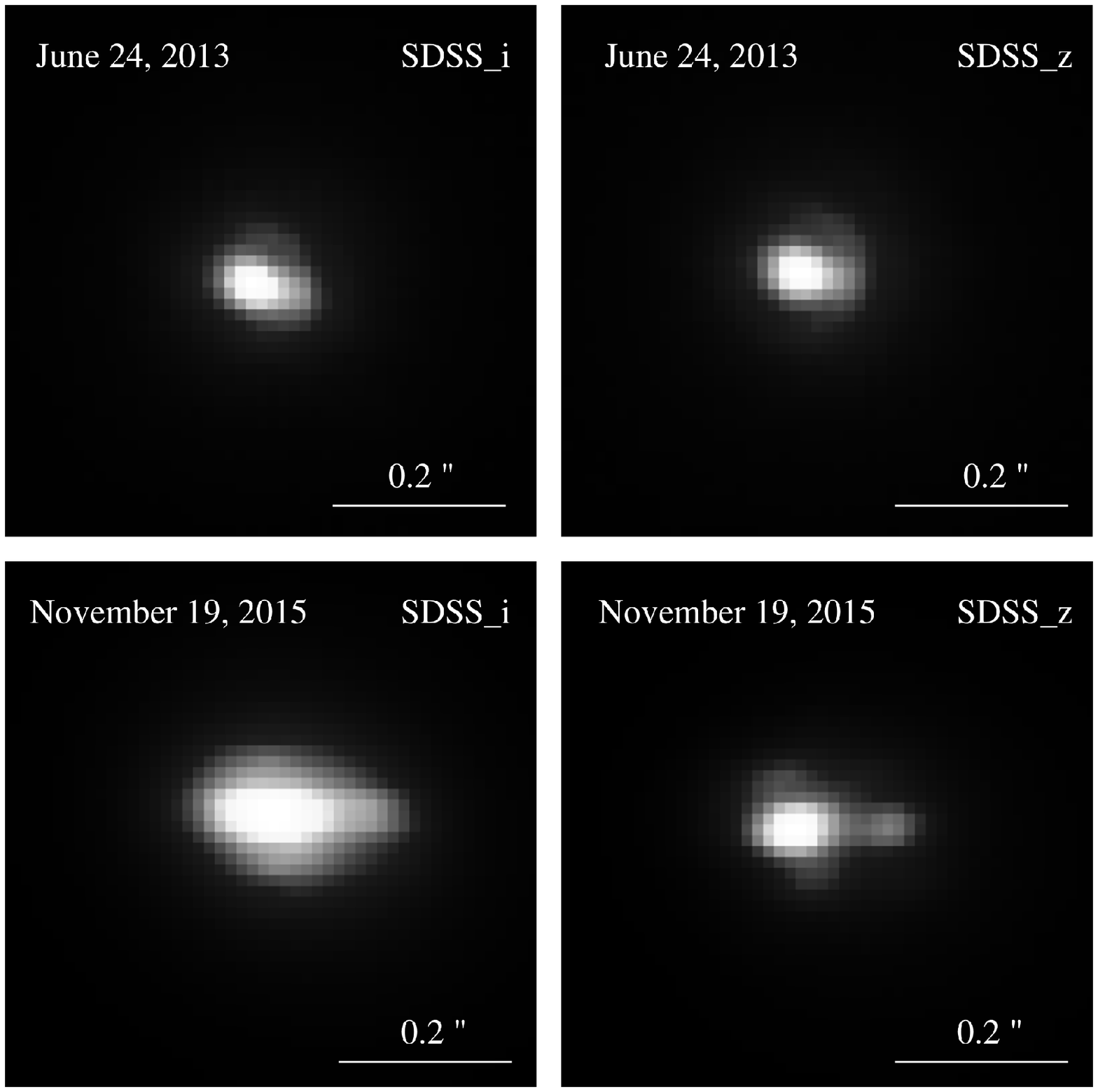}}
\caption[AstraLux images of the star HD\,160934]{AstraLux SDSS i$^\prime$ and SDSS z$^\prime$ images of the binary star HD\,160934 in 2013 and 2015; north is up and east is to the left. Both components of the star are not clearly distinguishable in the 2013 images but they are in the 2015 images.}
\label{hd_caha}
\end{center}
\end{figure*}

\begin{table*}
\caption{Binary properties of HD\,160934}            
\label{tablehd}
\begin{center}
%\resizebox{\hsize}{!}{
\begin{tabular}{c c c c c }     % 7 columns 
\hline\hline       
                       
Date & Separation & PA & Flux ratio & Reference\\    % table heading 
(Filter) & [$^{\prime\prime}$] & [$^\circ$] & & \\    % table heading
\hline
30 Jun 1998 & 0.155 $\pm$ 0.001 & 275.5 $\pm$ 0.2 & 0.485 $\pm$ 0.006 & Hormuth et al.~(2007)\\
(F165M) & & & & \\
17 Apr 2005 & 0.218 $\pm$ 0.002 & 268.5 $\pm$ 0.7 & 0.455 $\pm$ 0.021 & Lafreni\`{e}re et al.~(2007)\\
(NIRI$-$CH4) & & & & \\
8 Jul 2006 & 0.215 $\pm$ 0.002 & 270.9 $\pm$ 0.3 & 0.329 $\pm$ 0.051 & Hormuth et al.~(2007)\\
(RG830) & & & & \\
19 Nov 2015 & 0.23 $\pm$ 0.01 & 267 $\pm$ 1 & 0.5 $\pm$ 0.1 & This work\\
(SDSS z$^\prime$) & & & & \\
19 Nov 2015 & 0.21 $\pm$ 0.03 & 270 $\pm$ 5 & 0.6 $\pm$ 0.2 & This work\\
(SDSS i$^\prime$) & & & & \\
\hline                
\end{tabular}
%}
\end{center}
\footnotesize{\textbf{Notes.} The central wavelength of the filters are (in $\mu$m): 1.6511 (F165M), 
1.58 (NIRI-CH4), 0.910 (RG830), 0.907 (SDSS z$^\prime$), and 0.767 (SDSS i$^\prime$).}
\end{table*}

\begin{table*}
\caption{Compilation of astrometric measurements for the HD\,160934 system}             
\label{poss}      
\begin{center}
%\resizebox{\hsize}{!}{       
\begin{tabular}{c c r r c }
\hline\hline
% & & Relative positions AB\,Dor\,Ba $-$ AB\,Dor\,Bb & & \\       
\multicolumn{5}{c}{Relative positions of HD\,160934\,A $-$ HD\,160934\,c} \\       
                      
Epoch & Instrument & $\Delta\alpha$\,(mas)~~~~~~~~~ & $\Delta\delta$\,(mas)~~~~~~~ & Reference \\ 
\hline                    
1998.496 & HST/NICMOS (IR) & 154.3 $\pm$ 0.9~~~~~~~~~ & $-$14.8 $\pm$ 0.5~~~~~~~~~ & (1) \\
2005.296 & Gemini (IR) & 212.9 $\pm$ 2.0~~~~~~~~~ & 5.6 $\pm$ 2.6~~~~~~~~~ & (2) \\
2006.518 & AstraLux (IR) & 214.9 $\pm$ 1.0~~~~~~~~~ & $-$3.4 $\pm$ 1.0~~~~~~~~~ & (1) \\
2006.712 & Gemini (IR) & 217.9 $\pm$ 2.0~~~~~~~~~ & $-$4.9 $\pm$ 2.6~~~~~~~~~ & (2) \\
2008.477 & Palomar (IR) & $-$169.1 $\pm$ 0.3~~~~~~~~~ & $-$9.7 $\pm$ 0.3~~~~~~~~~ & (3) \\
2010.318 & Keck (IR) & 64.6 $\pm$ 0.3~~~~~~~~~ & $-$18.9 $\pm$ 0.3~~~~~~~~~ & (3) \\
2011.310 & Keck (IR) & $-$6.3 $\pm$ 0.3~~~~~~~~~ & $-$18.9 $\pm$ 0.3~~~~~~~~~ & (3) \\
2012.830 & EVN (radio) & $-$22.3 $\pm$ 0.1~~~~~~~~~ & 6.2 $\pm$ 0.2~~~~~~~~~ & (4) \\
2014.175 & EVN (radio) & 149.3 $\pm$ 0.1~~~~~~~~~ & 12.9 $\pm$ 0.2~~~~~~~~~ & (4) \\
2015.887 & AstraLux (IR) & 225.7 $\pm$ 3.0~~~~~~~~~ & 11.8 $\pm$ 3.0~~~~~~~~~ & (4) \\
\hline                  

\multicolumn{5}{c}{Absolute positions of HD\,160934 (EVN)} \\       

%% & & Absolute positions AB\,Dor\,B (LBA) & & \\

Epoch & Component & RA\,(h m s)~~~~~~~ & Dec\,($^{\circ}$\,$^\prime$\,$^{\prime\prime}$)~~~~~~~ & \\
\hline
2012.830 & A & 17 38 39.59830 $\pm$ 0.00014 & 61 14 16.6077 $\pm$ 0.0005 & (4)\\
 & c & 17 38 39.60138 $\pm$ 0.00014 & 61 14 16.6015 $\pm$ 0.0005 & (4)\\
2013.392 & A & 17 38 39.60667 $\pm$ 0.00016 & 61 14 16.6865 $\pm$ 0.0007 & (4)\\
2014.175 & A & 17 38 39.61159 $\pm$ 0.00013 & 61 14 16.6882 $\pm$ 0.0005 & (4)\\
 & c & 17 38 39.59090 $\pm$ 0.00013 & 61 14 16.6753 $\pm$ 0.0005 & (4)\\
\hline
\end{tabular}
%}
\end{center}
\footnotesize{\textbf{Notes.} (1) Hormuth et al.~(2007); (2) Lafreni\`{e}re et al.~(2007); (3) Evans et al.~(2012); (4) This work. The standard deviation of the relative position corresponds to the S/N-based uncertainty of the peaks of brightness of HD\,160934\,A and c. The absolute positions were obtained with reference to the IERS coordinate of the external quasar J1746+6226 ($\alpha=17^\mathrm{h}46^\mathrm{m}14^\mathrm{s}.034, \delta=62^\circ\,26^\prime\,54^{\prime\prime}.738$). The standard deviation of the absolute position includes, in addition to the uncertainty of the peak of brightness, the contribution of the propagation media and the reference source structure.}
\end{table*}

%This result allowed us to compare the luminosity of HD\,160934 with the luminosity of 
%something that had not been discovered so far. However, there is no evidence of 
%radio emission of the third low-mass component (HD\,160934B) supposed to be 
%at $\sim$8\farcs7 from the primary pair.
%Taking into account this, we could say that HD\,160934 would be like the AB\,Dor star in the northern hemisphere.

\subsubsection{Orbital parameters} 

% The orbital elements are the period (p), the time of periastron passage (T$_{0}$), the eccentricity (e), 
% the angle of line of nodes ($\Omega$), the inclination (i), the angle from node to periastron ($\omega$), 
% and the semimajor axis (a). The determined values for the orbit appears in Table \ref{table:2}.

In order to carry out an astrometric study, we measured the relative position of the pair A$-$c directly on the maps shown in Fig.~\ref{hd} (except for epoch 2013.39); we also measured the absolute position of the main component A, whose coordinates in Fig.~\ref{hd} are in turn referenced to the position of the external quasar (Table \ref{poss}). We augmented our data set with the relative position in Fig. \ref{hd_caha} of 2015 and with previous orbital measurements reported by Evans et al.~(2012), Hormuth et al.~(2007), and Lafreni\`{e}re et al.~(2007). Table \ref{poss} shows all the archive positions available for the system.

%\begin{table}
%\caption{HD\,160934\,c astrometry}
%\label{others}
%\centering
%\begin{tabular}{c c c c}
%\hline\hline
%Epoch & $\rho$ (mas) & $\theta$ (deg) & Source \\ 
%\hline
   %1998.49589 & 155 $\pm$ 1 & 275.5 $\pm$ 0.2 & Hormuth et al.~(2007) \\
   %2005.29589 & 213 $\pm$ 2 & 268.5 $\pm$ 0.7 & Lafreni\`{e}re et al.~(2007) \\
   %2006.51781 & 215 $\pm$ 2 & 270.9 $\pm$ 0.3 & Hormuth et al.~(2007) \\
   %2006.71233 & 218 $\pm$ 2 & 271.3 $\pm$ 0.7 & Lafreni\`{e}re et al.~(2007) \\
   %2008.47671 & 169.4 $\pm$ 0.3 & 273.3 $\pm$ 0.1 & Evans et al.~(2012) \\
   %2010.31781 & 68.8 $\pm$ 0.7 & 290.0 $\pm$ 0.6 & Evans et al.~(2012) \\
   %2011.30959 & 20.0 $\pm$ 0.1 & 18.43 $\pm$ 0.1 & Evans et al.~(2012) \\
%\hline
%\end{tabular}
%\end{table}

\begin{figure*}
\resizebox{\hsize}{!}
{\includegraphics{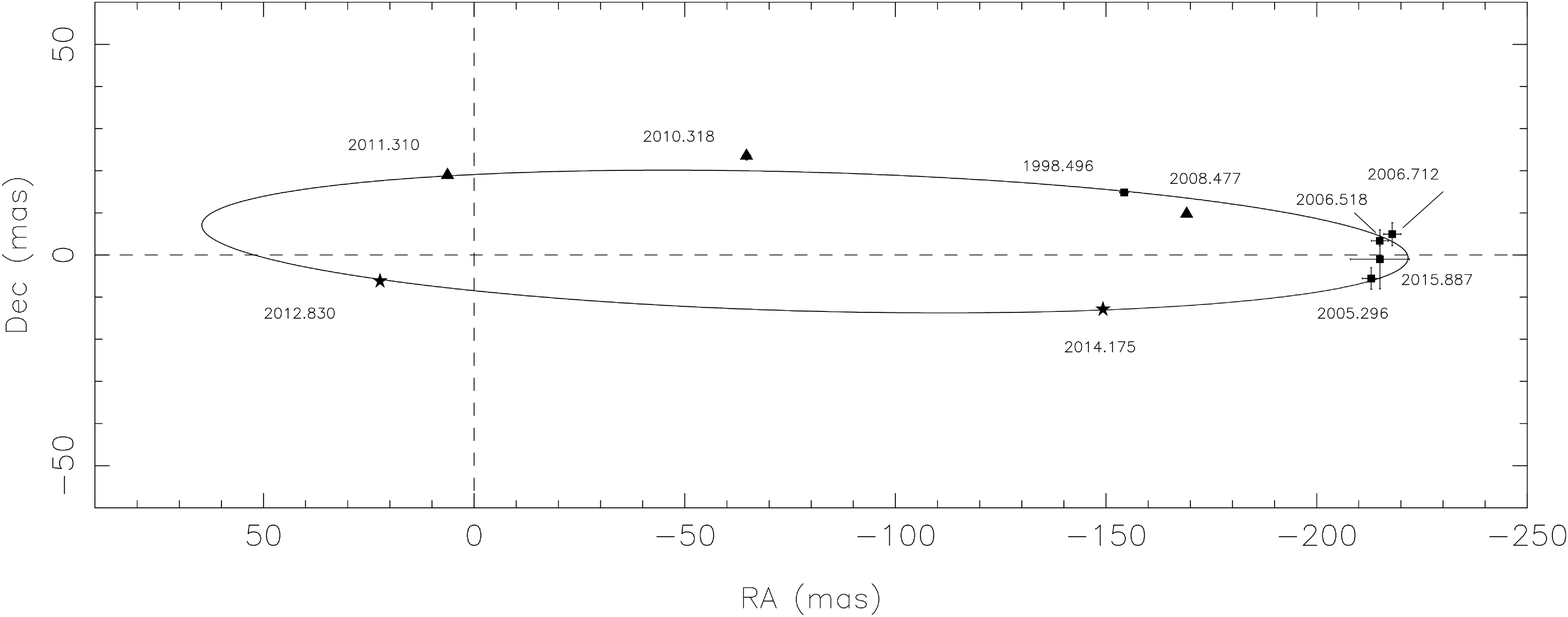}}
\caption[Relative orbit of the component HD\,160934\,c respect HD\,160934\,A]{Relative orbit of the binary star HD\,160934 using the orbital elements in Table \ref{tableorbitt} (with a$_{\mathrm{rel}}$). HD\,160934\,A is located at the origin of the figure, hence the relative orbit of HD\,160934\,c is depicted. Each type of symbol corresponds to a different technique to measure the relative position of HD\,160934\,c, namely, infrared relative astrometry (squares; Hormuth et al.~2007; Lafreni\`ere et al.~2007, this work), masking interferometry (triangles; Evans et al.~2012), and VLBI (star symbol; this work). Error bars are plotted but hardly visible because of the size of the orbit. For clarity, these are shown in Fig.~6.}
\label{orbit_relhd}
\end{figure*}

\begin{figure*}
\resizebox{\hsize}{!}
{\includegraphics{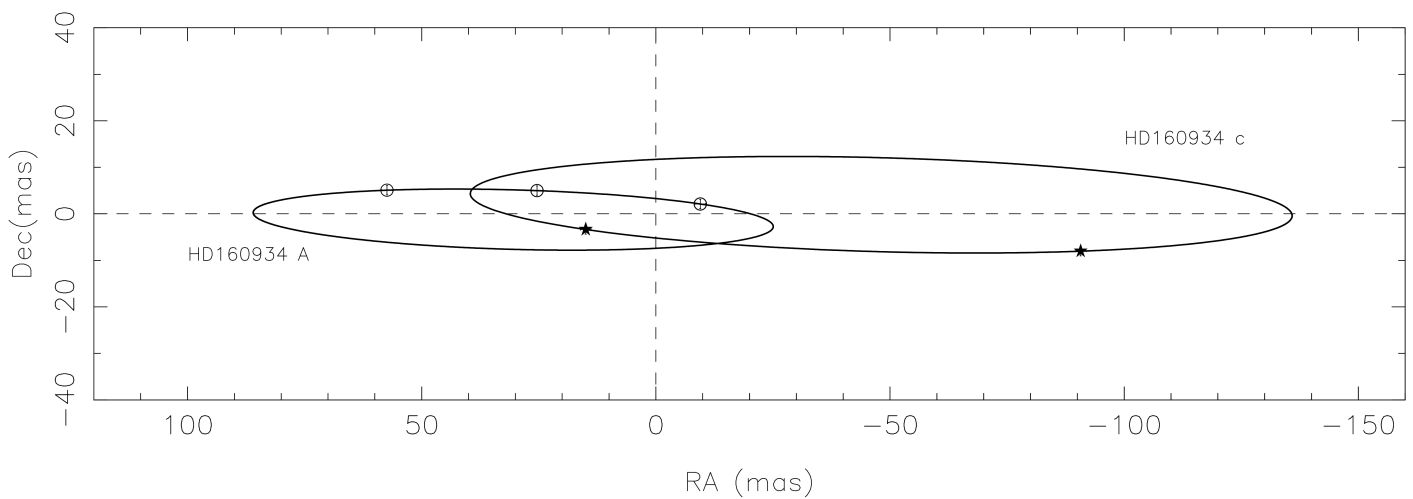}}
\caption[Absolute orbits of the components HD\,160934\,A and HD\,160934\,c]{Absolute orbits of the binary components HD\,160934\,A and HD\,160934\,c using the orbital elements in Table \ref{tableorbitt} (with a$_{\mathrm{A}}$ and a$_{\mathrm{c}}$, respectively). The positions of the component A (circles) and c (star symbols) are indicated. The center of mass of the system is placed at the origin.}
\label{orbit_abshd}
\end{figure*}

We estimated the Keplerian parameters of HD\,160934 via a weighted least-squares fit that combined the absolute positions of component A and all the relative positions constructed as A$-$c, that is, taking c as reference. We followed a similar approach to that used in Azulay et al. (2015) for another star of the AB\,Dor-MG, that is, AB\,Dor\,B,  solving simultaneously for the absolute and relative orbits using the Thiele-Innes elements and the Levenberg-Marquardt algorithm. In practice, we proceed with two steps:

\begin{enumerate} 
\item{We obtained \textit{a priori} values of the orbital elements from a previous least-squares fit to the, more numerous, differential data; in particular we estimated values for the period $P$ (10.33\,yr), semimajor axis of the relative orbit $a_{\mathrm{rel}}$ (0.$^{\prime\prime}$152), eccentricity $e$ (0.63), three orientation angles $i$ (82.$^{\circ}$4), $\omega$ (85.$^{\circ}$9), $\Omega$ (35$^{\circ}$), and time of periastron $T_{0}$ (2002.32).}
\item{We used the values above as (otherwise excellent) \textit{a priori} to favor the convergence of the L-M algorithm in the combined fit of the absolute (A component) and relative positions (A$-$c). In this analysis, the proper motion and parallax of the system were also estimated. The resulting set of astrometric and orbital parameters is shown in Table \ref{tableorbitt}, and meanwhile plots of the relative and absolute orbits are presented in Fig.~\ref{orbit_relhd} and \ref{orbit_abshd}.}
\end{enumerate}

Although there is a third object in this system, HD\,160934\,B (located at 8$^{\prime\prime}$ separation from A and c; Lowrance et al. 2005), we did not include secular acceleration terms in the fit described above, since this third body does not induce an appreciable acceleration in our three-year time baseline of VLBI monitoring; the estimated period of the corresponding reflex orbital motion is on the order of $10^3$ years). 

%%%%%%%%%%%%%%%%%%%%%%%%%%%%%%%%%
%%%%%%%%%%% CAMBIAR ESTA TABLA POR UNA SIMILAR A LA DEL AJUSTE DE ABDORB
%%%%%%%%%%%%%%%%%%%%%%%%%%%%%%%%%
\begin{table}
\begin{center}
\caption[Estimates of the astrometric and orbital parameters of HD\,160934]{Estimates of the astrometric and orbital parameters of HD\,160934$^\mathrm{a,b}$}
\label{tableorbitt}
\begin{tabular}{r l }
\hline\hline
Parameter & Value \\
\hline
$\alpha_{0}$\,(h m s): & 17 38 39.6349 $\pm$ 0.0002 \\
$\delta_{0}$\,($^\circ$ $^{\prime}$ $^{\prime\prime}$): & $+$61 14 16.0238 $\pm$ 0.0015 \\
$\mu_{\alpha}$\,(s yr$^{-1}$): & $-$0.0025 $\pm$ 0.0002 \\
$\mu_{\delta}$\,(arcsec yr$^{-1}$): & 0.0469 $\pm$ 0.0002 \\
%$Q_{\alpha}$\,(s yr$^{-2}$): & 0.000008 $\pm$ 0.000001 \\
%$Q_{\delta}$\,(arcsec yr$^{-2}$): & $-$0.00010 $\pm$ 0.00005 \\
$\pi$\,(arcsec)$^\mathrm{c}$: & 0.0314 $\pm$ 0.0005 \\
 & \\
$P$\,(yr): & 10.26 $\pm$ 0.08 \\
$a_{\mathrm{rel}}$\,($^{\prime\prime}$): & 0.1554 $\pm$ 0.0008 \\
$a_{\mathrm{A}}$\,($^{\prime\prime}$): & 0.0603 $\pm$ 0.0014\\
$a_{\mathrm{c}}$\,($^{\prime\prime}$): & 0.0952 $\pm$ 0.0014\\
$e$: & 0.64 $\pm$ 0.03 \\
$i\,(^\circ$): & 82.72 $\pm$ 0.12 \\
$\omega_{\mathrm{c}}\,(^\circ$)$^\mathrm{d}$: & 37.7 $\pm$ 0.5  \\
$\Omega\,(^\circ$): & 266.74 $\pm$ 0.12\\
$T_{0}$: & 2002.4 $\pm$ 0.1 \\
 & \\
$m_{\mathrm{A}}$\,(M$_{\odot}$): & 0.70 $\pm$ 0.07\\ 
$m_{\mathrm{c}}$\,(M$_{\odot}$): & 0.45 $\pm$ 0.04\\ 
\hline
\end{tabular}
\end{center}
\footnotesize{\textbf{Notes.} $^\mathrm{a}$ The reference epoch is 2000.0.
$^\mathrm{b}$ The number of degrees of freedom of the fit is 14; the minimum value found for 
the reduced $\chi^2$ is 1.2.
$^\mathrm{c}$ Our parallax estimate is more accurate than the \textit{Hipparcos} value given for AB\,Dor\,A but fully compatible with that within the uncertainty.
$^\mathrm{d}$ For the absolute orbit of HD\,160934\,A, $\omega_\mathrm{A}=\omega_\mathrm{c}+\pi$.}
\end{table}

%\begin{table*}
%\caption{Orbital parameters of HD\,160934}
%\label{tableorbitt}
%\begin{center}
%\begin{tabular}{l r@{\,$\pm$\,}l }
%\hline\hline
%Parameter & \multicolumn{2}{c}{Value}\\
%\hline
   %$P$\,(yr): & 10.33  &  0.06 \\
   %$a$\,("): & 0.152  &   0.002 \\
   %$e$: & 0.626  &   0.005 \\
   %$i\,(^\circ$): & 82.4  &   0.2 \\
   %$\omega\,(^\circ$): & 35 &   1 \\
   %$\Omega\,(^\circ$): & 85.9 &   0.3 \\
   %$T_{0}$: & 2002.32  &   0.07 \\
%\hline
%\end{tabular}
%\end{center}
%\footnotesize{\textbf{Notes.} The uncertainties shown are the standard errors scaled so that the $\chi^2$ per degree of freedom is unity.}
%\end{table*}

\begin{figure}
\begin{center}
\resizebox{0.5\textwidth}{!}
{\includegraphics{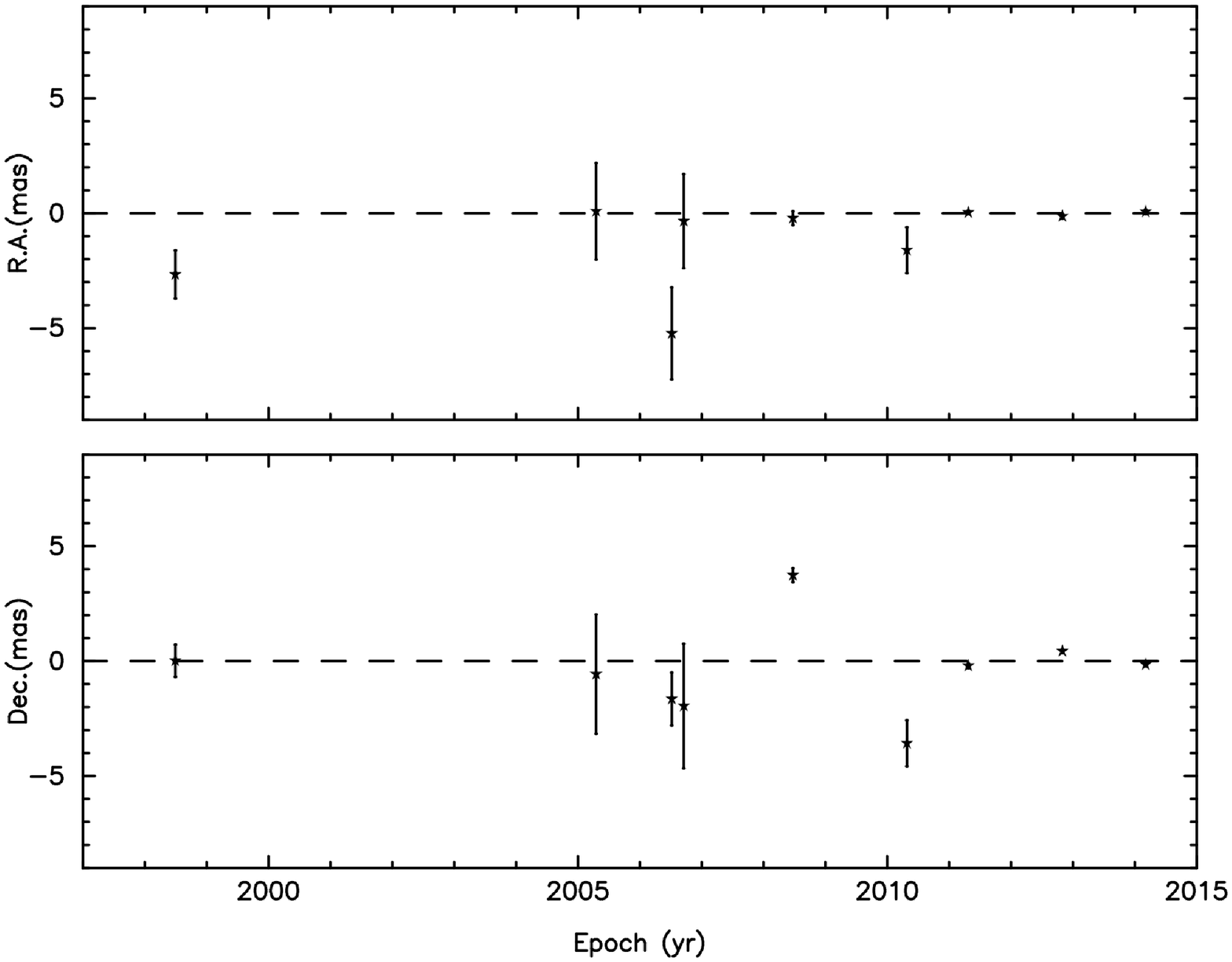}}
\caption[Residuals in right ascension and declination of the relative orbit]{Postfit residuals in right ascension (\textit{upper panel}) and declination (\textit{lower panel}). The weighted rms of the plotted residuals is 3.2\,mas.}
\label{postfit_hd}
\end{center}
\end{figure}

Our fit yielded a new value of the parallax (31.4$\pm$0.5\,mas), which is within the uncertainties, but more precise than the previous \textit{Hipparcos} estimate (30.2$\pm$2\,mas; van Leeuwen 2007). This new parallax allowed us to determine the sum of the masses of both components of HD\,160934 using Kepler's third law
\begin{eqnarray*}
\frac{(a_{rel}^{\prime\prime} / \pi^{\prime\prime})^{3}}{P^{2}} = (m_{\mathrm{A}} + m_{\mathrm{c}})_{\odot} \mbox{.}
\end{eqnarray*}
We obtained a value of $m_{\mathrm{A}} + m_{\mathrm{c}}$ of 1.15$\pm$0.10 M$_\odot$, which is coincident with previous estimates made by other authors (Evans et al.~2012 and references therein). Similarly, using the semimajor axis of the absolute orbit of component A, $a_{\mathrm{A}}$, we could estimate the mass of component c, $m_{\mathrm{c}}$, using $m_{\mathrm{c}}^3 / (m_{\mathrm{A}} + m_{\mathrm{c}})^2 = a_{A}^3/P^2$, which yielded a value of 0.45$\pm$0.04\,M$_\odot$. A value of the $m_{\mathrm{A}}$ (= 0.70$\pm$0.07\,M$_{\odot}$) follows from a simple subtraction of the values above. In principle, the latter value for $m_{\mathrm{A}}$ may seem highly correlated with $\mathrm{c}$; however, a similar, but coarser, estimate of $a_{\mathrm{c}}$ could be obtained by repeating the combined fit using the absolute positions of component c (see Table \ref{poss}), from which a similar value of $m_{\mathrm{A}}$ (0.70$\pm$0.10\,M$_{\odot}$) could be calculated. The coincidence of both estimates of $m_{\mathrm{A}}$ indicates the robustness of our mass determinations.

The weighted rms of the postfit residuals (plotted in Fig.~\ref{postfit_hd}) is 3.2\,mas, meaning that some unmodeled effects are still present in the data. The residuals show no evidence of another companion within the errors. Instead, the possible departure of some of the points from the fitted orbit might indicate instrumental effects that have not been considered. Accordingly, we scaled the statistical errors of the orbital parameters to take this contribution into account (see Table \ref{tableorbitt}).

\subsubsection{Comparison with models}
To proceed with the calibration of PMS models we needed the K magnitude and effective temperature of components A and c. Estimates of the individual K magnitudes can be obtained from the unresolved JHK 2MASS photometry, that is 6.812 $\pm$ 0.020, combined with the flux ratio between c and A ($f_{\mathrm{c/A}}$) at K band; however, although $f_{\mathrm{c/A}}$ has been measured for different filters (see Table \ref{tablehd}), the K-band magnitude is not available. We estimated $f_{\mathrm{c/A}}$ at K band via weighted linear fits to the flux ratios measured at other filters given in Table \ref{tablehd}. We obtained a value of $f_{\mathrm{c/A}} = 0.39 \pm 0.11$ as a mean of our fits, and the standard deviation as the spread in the fits considering different subsets of data points, in such a way that our uncertainty conservatively covers the different fitted values of $f_{\mathrm{c/A}}$. Final values for the absolute K magnitudes were 4.65$\pm$0.15 and 5.68$\pm$0.15 for components A and c, respectively.

Regarding the effective temperatures, we derived a value for component A from its spectral type (K7$-$K8; McCarthy \& White 2012) using the empirical color-temperature transformation reported by Hartigan et al.~(1994). For component c, we proceeded in a similar way by assuming a M2$-$3 spectral type (G\'alvez et al.~2006). The final values for the effective temperatures were 3960$\pm$60\,K and 3450$\pm$90\,K for components A and c, respectively. The standard deviation of these temperatures were estimated to cover several uncertainties that could affect our approach. Recent studies have shown that the spectral types of young stars determined from NIR observations disagree with those from the optical by up to three subtypes (Kastner et al. 2015; Pecaut et al. 2016), which would strongly affect the procedure to estimate the effective temperature. Since the spectral types of HD\,160934 A and c are based on IR observations, we should not expect such large errors. Actually, the spectral type determination used for the components of HD\,160934 (see above) are 1 subtype uncertain, which has been taken into account to estimate the standard deviation of the effective temperatures. Considering that these standard deviations have been further enlarged to cope with the different, more recent, color-temperature relations (Pecaut \& Mamajek 2013; Luhman et al. 2003; Herczeg \& Hillenbrand 2014) this wavelength-dependent spectral type effect should be reasonably covered.

\begin{figure*}
\begin{center}
\resizebox{\hsize}{!}
{\includegraphics{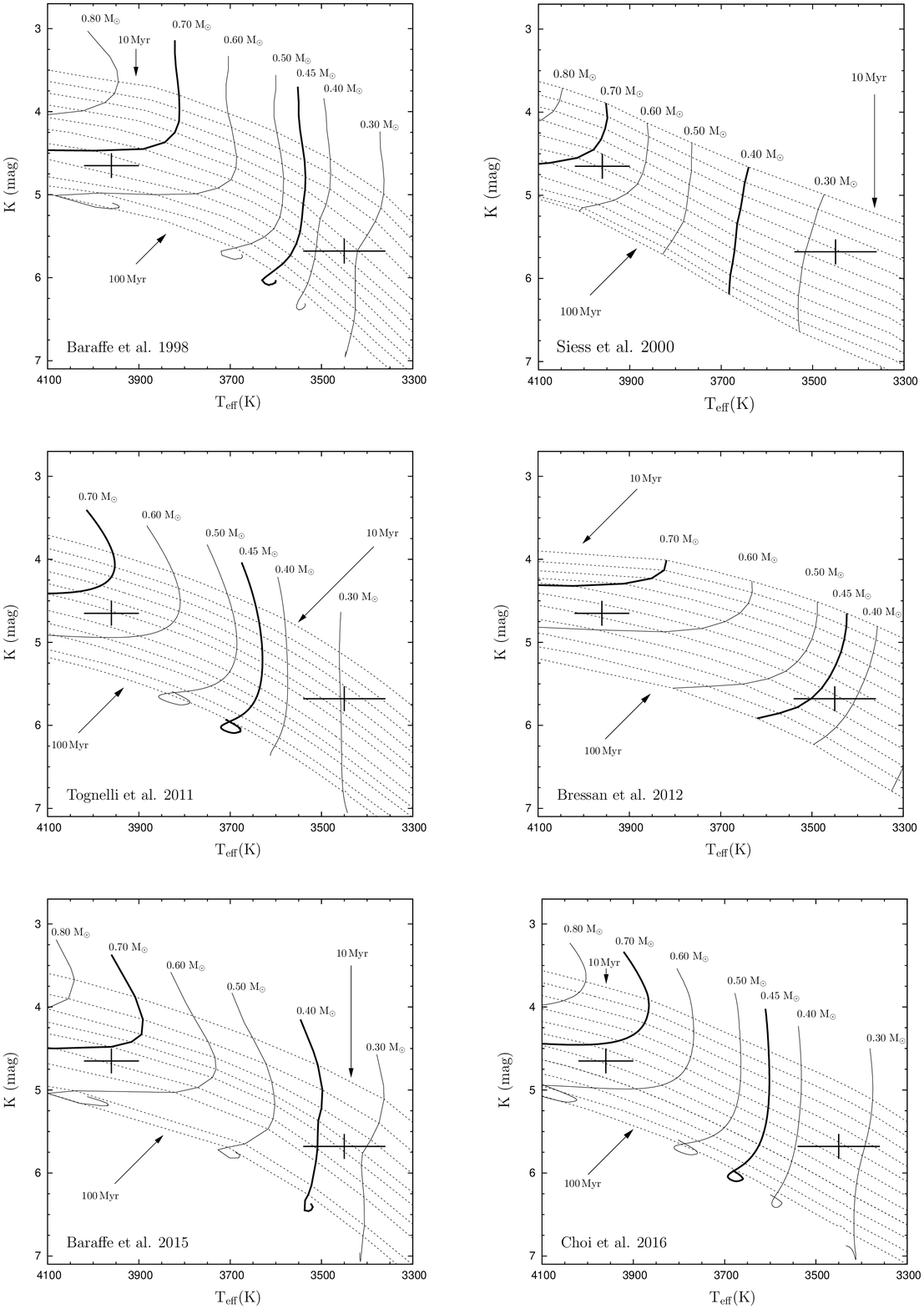}}
\caption[Comparison of HD\,160934 components with some PMS theoretical models]{Comparison of HD\,160934 components with some PMS theoretical models (Baraffe et al.~1998, \textit{top left}; Siess et al.~2000, \textit{top right}; Tognelli et al.~2011, \textit{middle left}; and Bressan et al.~2012, \textit{middle right}; Baraffe et al.~2015, \textit{bottom left}; and Choi et al.~2016, \textit{bottom right}). For each model, isomasses (solid lines) and isochrones (dashed lines) are plotted. We highlight the nearest tracks available corresponding to our dynamical mass values. The theoretical masses are consistent with our dynamical estimates just at the extreme of their uncertainties.}
\label{mod_hd}
\end{center}
\end{figure*}

In order to calibrate the stellar evolution models for PMS stars, we considered isochrones and isomasses corresponding to the models of Baraffe et al. (1998; BCAH98), Siess et al. (2000; S00), Tognelli et al. (2011; 2012; TDP12), Bressan et al. (2012; Padova), Baraffe et al. (2015; BHAC15), and Choi et al. (2016; MIST). We adopted a metallicity value of [Fe/H]$=$0.0 (representative of the AB\,Dor metallicity; Barenfeld et al. 2013) and a solar calibrated mixing length parameter $\alpha$. The different models can be shown in Fig.~\ref{mod_hd}. HD\,160934\,A and HD\,160934\,c are placed in the H-R diagrams in Fig.~\ref{mod_hd} using the values of both the magnitudes and temperatures explained above.

The theoretical masses predicted by the models agree with our dynamical estimates just at the extreme of their uncertainties. All sets of tracks predict masses for component A $\sim$10\% lower than our dynamical values, while predictions for the component c vary according to the model: BCAH98, BHAC15, and MIST predict masses that are $\sim$30\% lower, S00 and TDP12 predict masses that are $\sim$40\% lower, and Padova predict masses that are $\sim$10\% lower. These results are consistent with previously published works, which conclude that PMS stellar evolution models for low-mass stars underestimate the dynamical values between 10$-$30\% (Hillenbrand \& White 2004; Mathieu et al. 2007). According to the values above, predictions are better for component A than for component c, that is, the larger the dynamical mass, the smaller the difference between the theoretical and dynamical estimates.

\begin{figure*}
\begin{center}
\resizebox{\hsize}{!}
{\includegraphics{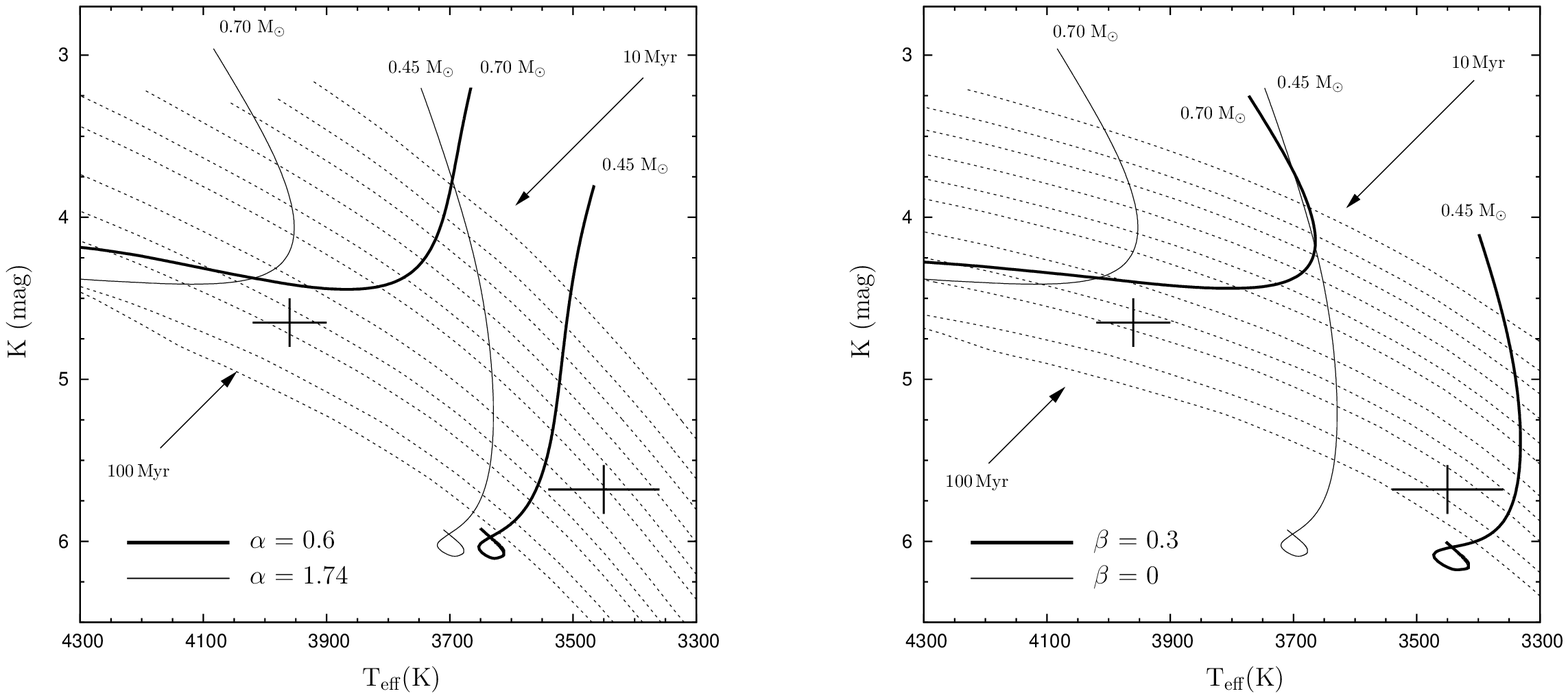}}
\caption[]{TPD12 models recalculated to include the effects of the stellar magnetic field. \textit{Left}: effect of an internal magnetic field, simulated by using a value of $\alpha$ = 0.6. The isomasses of the solar value $\alpha$ = 1.74 are also plotted for comparison. \textit{Right}: effect of stellar spots, computed for an effective spot coverage of $\beta$ = 0.3 (Somers \& Pinsonneault 2015). The isomasses for $\beta$ = 0 (standard models with no spots) are also plotted for comparison. In both cases, the isochrones (dashed lines) of the altered models are also shown and are to be compared with the isochrones of the TPD12 models in Fig. 7. See text.}
\label{moremod}
\end{center}
\end{figure*}

In terms of age, the S00 and TDP12 models suggest that both stars are younger than 40\,Myr, meanwhile BCAH98, Padova, BAHC15, and MIST favor slightly older ages but younger than 60\,Myr. The age of the HD\,160934 system predicted by the models in Fig.~\ref{mod_hd} is between 15\,Myr and 60\,Myr, near the young estimates of the age of the AB\,Dor-MG. Age estimates of the AB\,Dor-MG range from 50$-$120\,Myr (Malo et al. 2013), 70\,Myr (Zuckerman et al. 2011), 70$-$120\,Myr (Gagn\'e et al. 2014), 100$-$150\,Myr (Elliott et al. 2016), and 130$-$200\,Myr (Bell et al. 2015). It should be mentioned that the possibility that HD\,160934 does not belong to the AB\,Dor-MG cannot be ruled out. According to the BANYAN II membership probability tool (Gagn\'e et al. 2014), the kinematics of this object favor HD\,160934 to be a young field dwarf with a 95\% probability. Were this the case for HD\,160934, the conclusions about the age of the AB\,Dor-MG based on the age range derived from Fig. 7 would have a limited validity. However, in contrast with the BANYAN prediction, a very recent publication (Elliott et al. 2016) still confirms HD\,160934 as an AB\,Dor-MG member.

\subsubsection{Magnetic field effects}

The discrepancy between the dynamical and the inferred mass for the components of HD\,160934 seems to show the lack of additional input physics in the models. Indeed, the models appear to be too hot for the data at a given mass. To this regard, several hypothesis exist that would allow the models to be cooler; in the case of HD\,160934, and given the existence of compact radio emission on both objects, which is frequently associated with intense magnetic activity (i.e., G\"udel et al. 1995), we limited our analysis to study the impact of a magnetic field in the stellar models.

The presence of a magnetic field inside the star mainly acts to reduce the surface convection efficiency in stellar models, as shown in Feiden et al. (2012, 2013, 2014, 2016), thus reducing the stellar effective temperature, and increasing the star's radius. Moreover, an external magnetic field might also result in surface spots that, blocking the flux at the stellar surface, tend to increase the stellar radius (Somers \& Pinsonneault 2015). Thus, we briefly analyze these two aspects in turn.

Regarding the internal magnetic field and following Feiden et al. (2013), its main effect on the convective heat transport can be partially simulated by reducing the efficiency of super-adiabatic convection, which is effectively carried out using a value of the mixing-length parameter $\alpha$ that is much lower than the solar calibrated value. Consequently, we computed new TDP12 models using a value of $\alpha$ = 0.6, in contrast with the solar calibrated $\alpha$ = 1.74 value used in the models of Fig. 7. This particular value of $\alpha$ has been adopted by Feiden et al. (2013) to simulate the reduced convection efficiency due to a magnetic field in a non-magnetic stellar model; also such a low $\alpha$ value is compatible with those used by Chabrier et al. 2007 to reproduce the radius of low-mass eclipsing binaries stars. The results are shown in Fig. \ref{moremod} (\textit{left}), where we see an evident ``cooling'' of the magnetic ($\alpha$ = 0.6) isomasses of both components A and c with respect to those corresponding to the standard models ($\alpha$ = 1.74; see Fig. 7), which in turn produces a better agreement with the measurements. Interestingly, the effect of the internal magnetic field leads to older ages for both A ($>$50\,Myr) and c (>30\,Myr) components.

Concerning the effect of surface spot coverage on the models, we computed additional TDP12 evolutionary models, in which we implemented this effect following the formalism described in Somers \& Pinsonneault (2015). We adopted two different values of the effective spot coverage $\beta$ of 0 (standard models without spots; Fig. 7) and 0.3, which is the same value used by Somers \& Pinsonneault (2015). The comparison can be seen in Fig. \ref{moremod} (\textit{right}), where we see similar effects to those shown for the internal magnetic field: cooling of the isomasses, leading to a reduction of the discrepancy with the measurements, and older ages ($>$50\,Myr) for both components.

There are other possible scenarios that can modify the position of a pre-MS star in the HR diagram, such as the presence of protostellar accretion, but its treatment is beyond the scope of this paper; in such a scenario, it is difficult to define a standard set of accretion models, as the outputs of the models are severely affected by the parameters that govern the accretion phase (e.g., mass accretion rate, initial seed mass, or thermal energy carried inside the star by the accreted matter; Baraffe et al. 2012; Tognelli et al. 2015).

%%%fr(z)= 0.47 
%%%fr(i)= 0.55 

%\begin{table}
%\caption{Unresolved 2MASS photometry of HD160934}            
%\label{unresolvedhd}
%\begin{center}
%\begin{tabular}{c c }     % 7 columns 
%\hline\hline       
                       
%Filter/Color & Magnitude \\    % table heading 
%\hline
%J & 7.618 $\pm$ 0.024 \\
%H & 6.998 $\pm$ 0.016 \\
%K & 6.812 $\pm$ 0.020 \\
%\hline                
%\end{tabular}
%\end{center}
%\end{table}

\subsection{EK\,Draconis}
EK\,Dra (=HD\,129333) is an active, G1.5\,V star with a rapid rotation (2.6 days) (J\"arvinen et al.~2005). The binarity of this object (whose components are EK\,Dra\,A/B, separated by 0.74$^{\prime\prime}$) was discovered for the first time through radial velocity variations by Duquennoy \& Mayor (1991). Metchev \& Hillenbrand (2004) confirmed the existence of these components from IR imaging. Several radial velocity studies of this star have been carried out (Duquennoy \& Mayor 1991, Dorren \& Guinan 1994, Montes et al.~2001, K\"onig et al.~2005). In particular, K\"oning et al.~(2005) combined these radial velocity data with their data of speckle interferometry to derive masses of 0.9$\pm$0.1\,M$_{\odot}$ and 0.5$\pm$0.1\,M$_{\odot}$, for the primary and secondary, respectively, a period of 45$\pm$5\,yr, and a semimajor axis of 14.0$\pm$0.5\,AU.
%Moreover, EK\,Dra\,A is a known radio emitter with reported flux density at VLA scales of 0.35\,mJy (G\"udel et al.~1995).

The VLA image (Fig.~\ref{ekdra_vla}) shows EK\,Dra as an unresolved radio emitter with an integrated flux of 0.21\,mJy (the radio emission of EK\,Dra\,A at radio wavelengths is known and reported in G\"udel et al.~1995). Because of the small separation of both components of the binary at the epoch of observation, the components A and B appear to be blended on the map. Besides this, we could only detect the component A in the first of our three VLBI epochs of observations (2012.827). The image yields a flux density of 0.06\,mJy, with an upper bound to the radio emission of the component B of 0.02\,mJy (Fig.~\ref{ekdra}). The upper bounds to the radio emissions of the star in the second and third epochs are 0.01 and 0.02\,mJy, respectively. The non-detection of EK\,Dra in the last two VLBI epochs can be a consequence of the variable behavior of the radio emission. In the AstraLux images (Fig.~\ref{ekdra_caha}), meanwhile, we could detect both components of the star.

%EK\,Dra (=HD\,129333) is an active, G1.5\,V star with a rapid rotation (2.6 days) (J\"arvinen et al. 2005). The binarity of this star (which components are EK\,Dra\,A/B, separated 0.74") was discovered for the first time through radial velocity variations by Duquennoy \& Mayor (1991). Metchev \& Hillenbrand (2004) confirmed the existence of these components from IR imaging.

%Several radial velocity studies of this star have been carried out (Duquennoy \& Mayor 1991, Dorren \& Guinan 1994, Montes et al. 2001a, K\"oning et al. 2005). In particular, K\"oning et al. 2005 combined these radial velocity data with his data of speckle interferometry to derive masses of 0.9$\pm$0.1\,M$_{\odot}$ and 0.5$\pm$0.1\,M$_{\odot}$, for the primary and secondary, respectively, a period of 45$\pm$5\,yr and a semimajor axis of . 14$\pm$0.5\,AU. Moreover, EK\,Dra\,A is a known radio emitter with reported flux density at VLA scales of 0.35\,mJy (G\"udel et al. 1995).

%In our three epochs of observations, we could only detect the component A in the first one (with an upper bound to the radio emission of the B component of 0.02\,mJy). The upper bounds to the radio emissions of the star in the second and third epochs are 0.01 and 0.02\,mJy, respectively.

\begin{figure*}
\begin{center}
\resizebox{\hsize}{!}
{\includegraphics{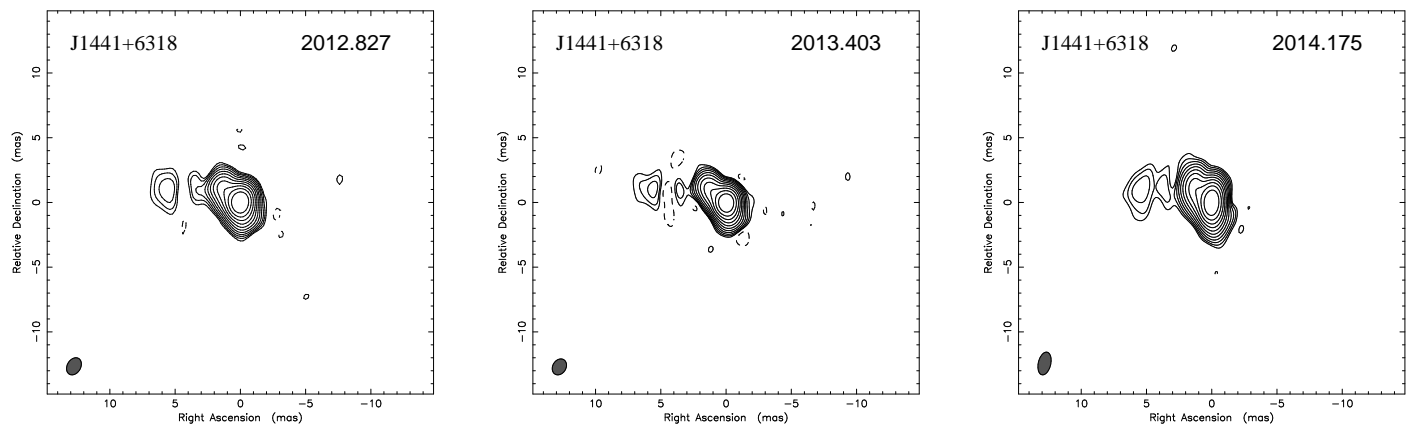}}
\caption[EVN maps of the calibrator J1441$+$6318]{EVN 5\,GHz images of J1441$+$6318 (calibrator of EK\,Dra) taken on 2012.827, 2013.403, and 2014.175, respectively. In each map, the lowest contour level corresponds to 5 times the statistical rms noise (0.3, 0.4, and 0.4\,mJy\,beam$^{-1}$) with a scale factor between contiguous contours of $\sqrt{3}$. The peak flux densities in the images are 0.17, 0.17, and 0.19 \,mJy\,beam$^{-1}$, respectively.}
\label{j1441}
\end{center}
\end{figure*}

\begin{figure}
\begin{center}
\resizebox{0.8\hsize}{!}
{\includegraphics{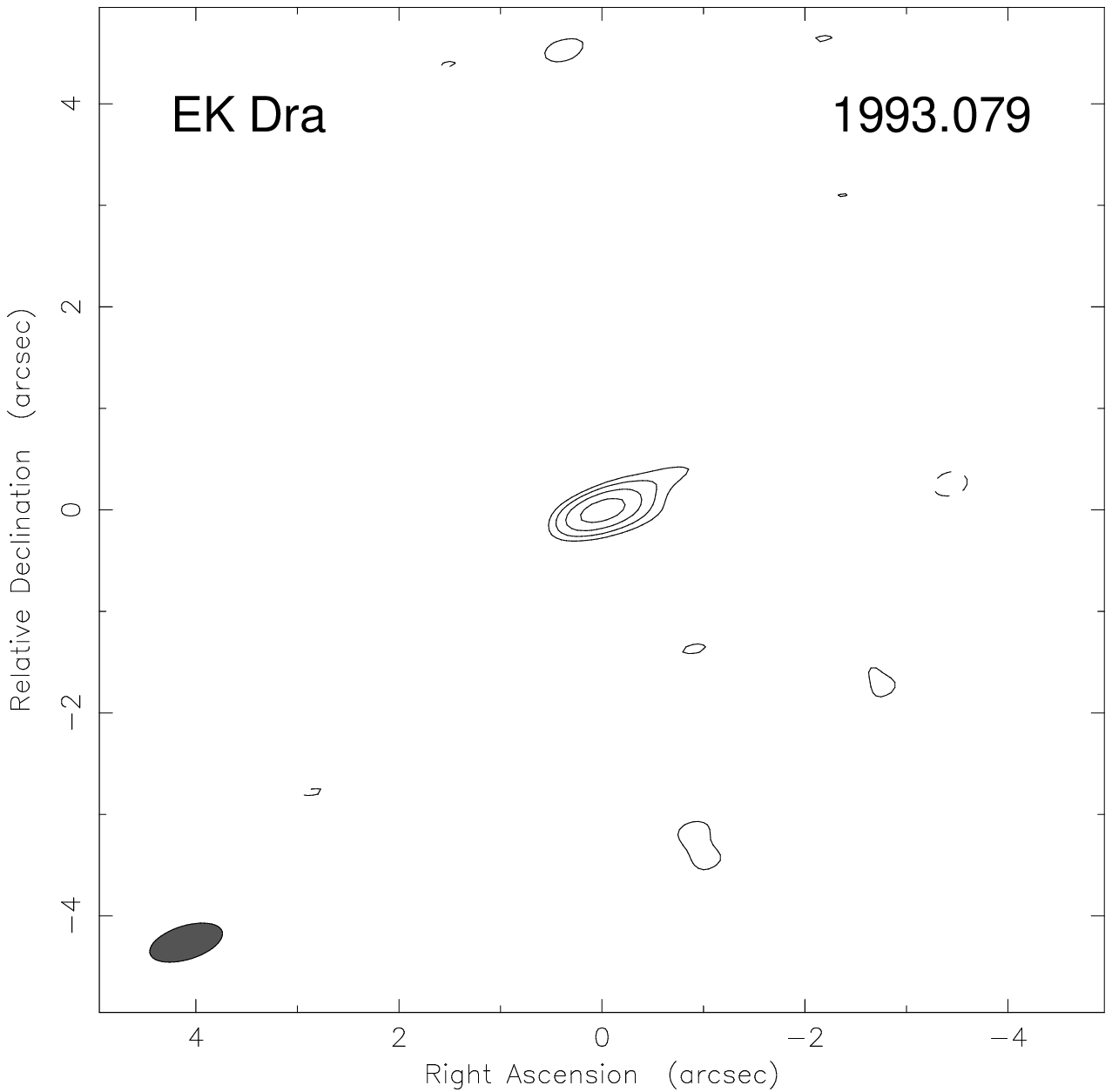}}
\caption[VLA map of the star EK\,Dra]{VLA 8.4\,GHz image of EK\,Dra taken on 1993.079. The lowest contour level corresponds to twice the statistical rms noise (0.02\,mJy\,beam$^{-1}$) with a scale factor between contiguous contours of $\sqrt{2}$. The peak flux density in the image is 0.18\,mJy/beam. The restoring beam (shown in the bottom-left corner) is an elliptical Gaussian of $0.74\times0.33$\,arcsec (PA $-$72.$^{\circ}$8).}
\label{ekdra_vla}
\end{center}
\end{figure}

\begin{figure}
\begin{center}
\resizebox{0.8\hsize}{!}
{\includegraphics{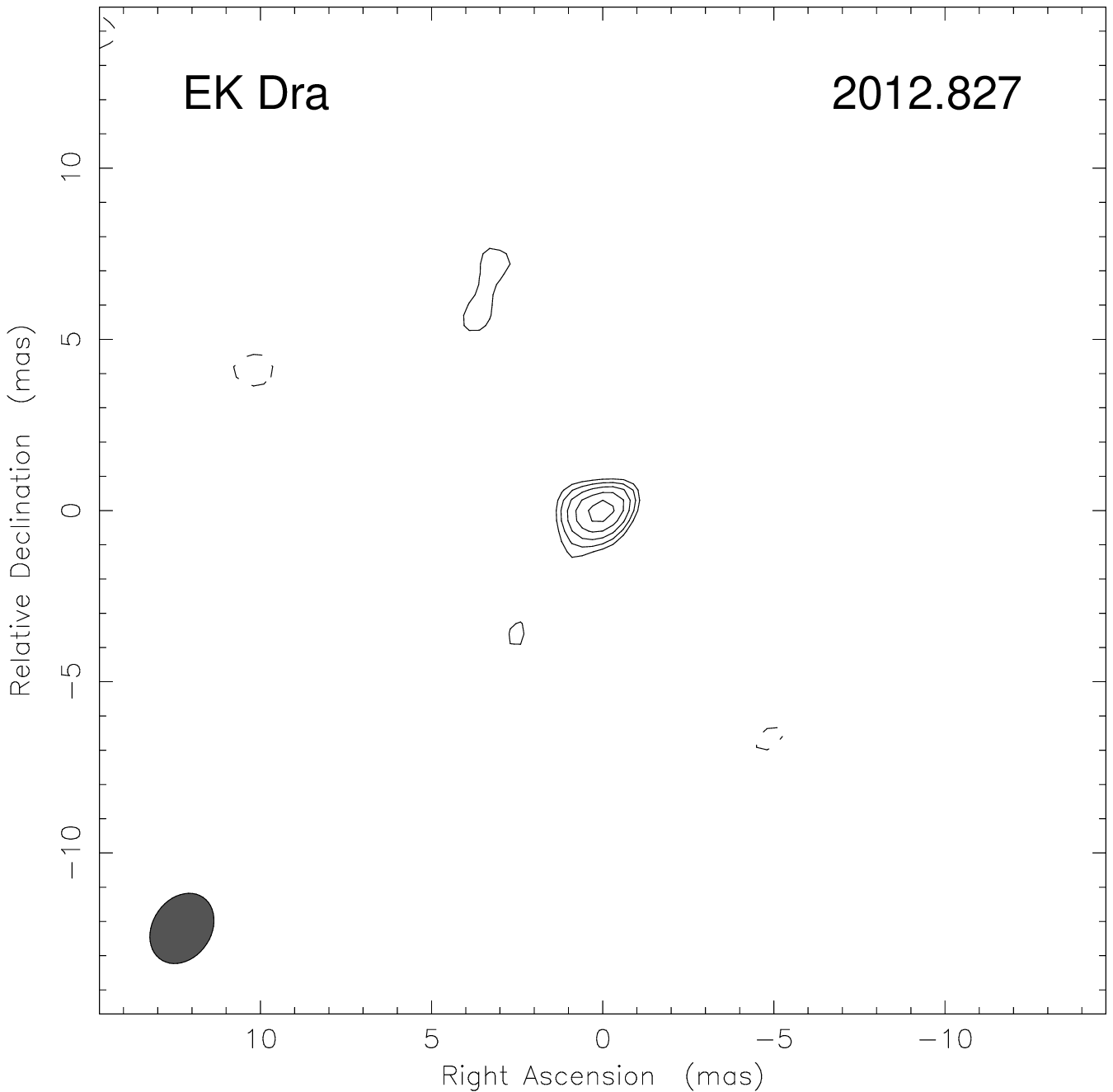}}
\caption[EVN maps of the star EK\,Dra]{EVN 5\,GHz image of EK\,Dra taken on 2012.827. The lowest contour level corresponds to 3 times the statistical rms noise (0.02\,mJy\,beam$^{-1}$) with a scale factor between contiguous contours of $\sqrt{2}$. The peak flux density in the image is 0.06\,mJy\,beam$^{-1}$. The restoring beam (shown in the bottom-left corner) is an elliptical Gaussian of $2.19\times1.71$\,mas (PA $-$34.$^{\circ}$1).}
\label{ekdra}
\end{center}
\end{figure}

\begin{figure*}
\begin{center}
\resizebox{0.6\hsize}{!}
{\includegraphics{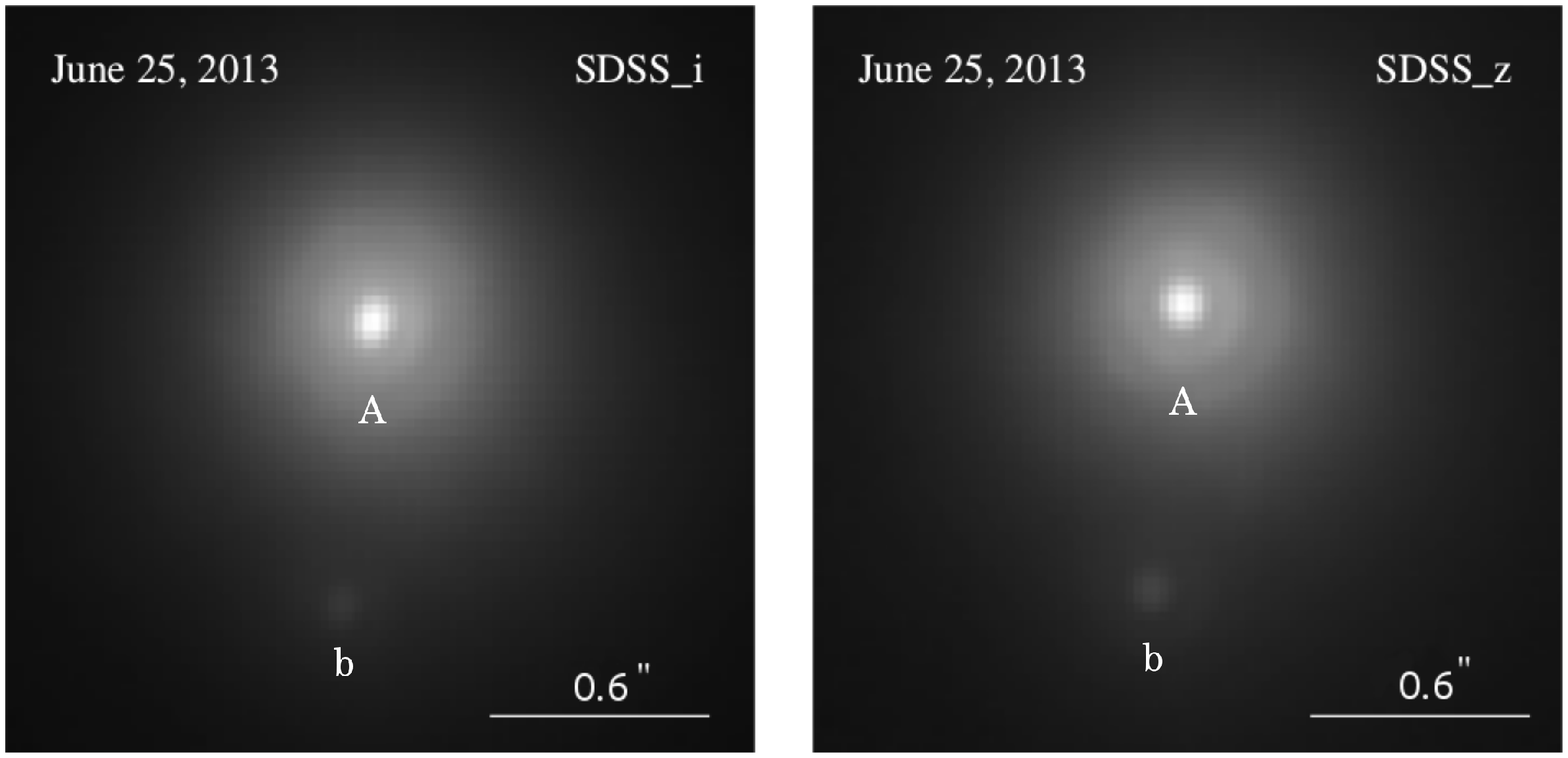}}
\caption[AstraLux maps of the star EK\,Dra]{AstraLux SDSS\_i and SDSS\_z images of the binary star EK\,Dra in 2013; north is up and east is to the left. Both components of the star, component A (north) and B (south), are clearly detectable.}
\label{ekdra_caha}
\end{center}
\end{figure*}

\begin{figure}
\begin{center}
\resizebox{0.6\hsize}{!}
{\includegraphics{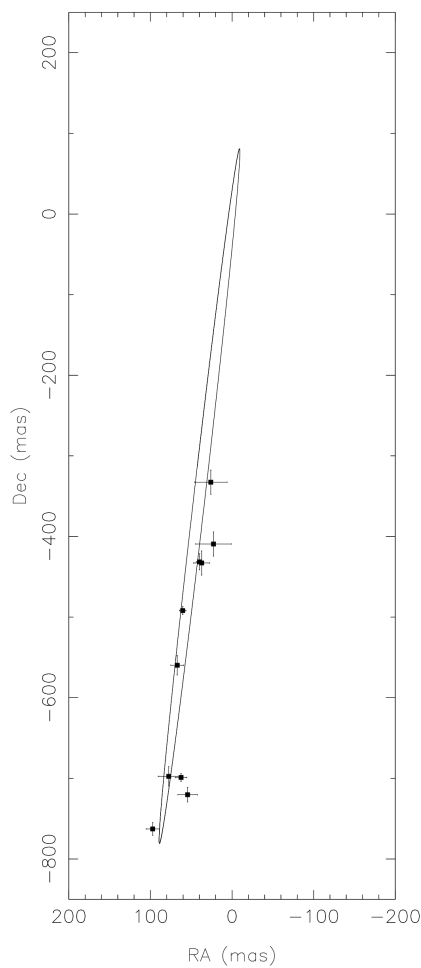}}
\caption[Relative orbit of the component EK\,Dra\,B respect EK\,Dra\,A]{Relative orbit for the binary EK\,Dra using the orbital elements in Table \ref{tableorbitt_ek}. EK\,Dra\,A component is located at the origin.}
\label{o1_ek}
\end{center}
\end{figure}

\begin{figure}
\begin{center}
\resizebox{0.8\hsize}{!}
{\includegraphics{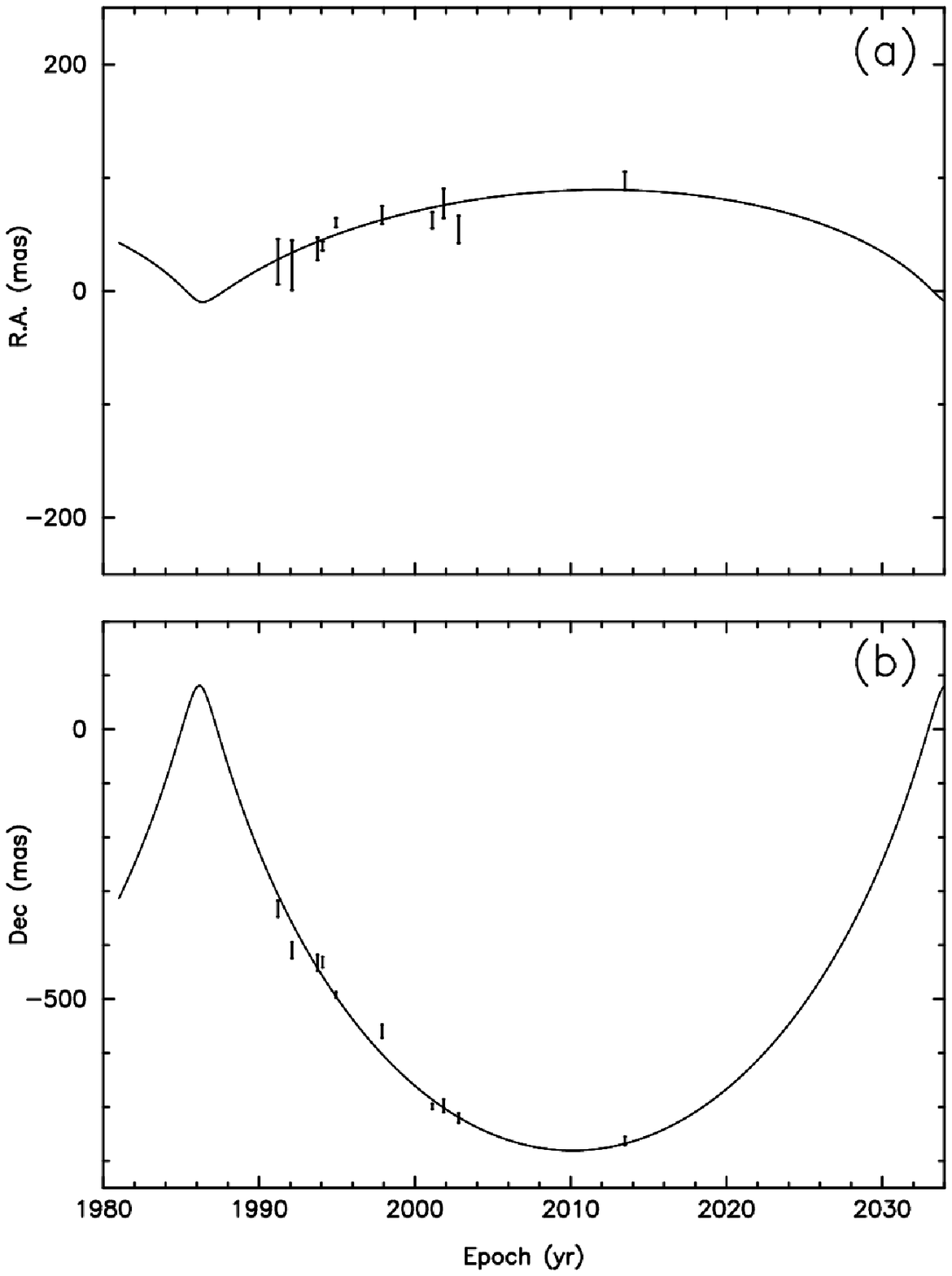}}
\caption[Relative orbit of EK\,Dra in RA and Dec]{Orbital motion of the binary EK\,Dra in right ascension (a) and declination (b). The line corresponds to the least-squares fitted values of Table \ref{tableorbitt_ek}.}
\label{o2_ek}
\end{center}
\end{figure}

\begin{figure}[h]
\begin{center}
\resizebox{0.8\hsize}{!}
{\includegraphics{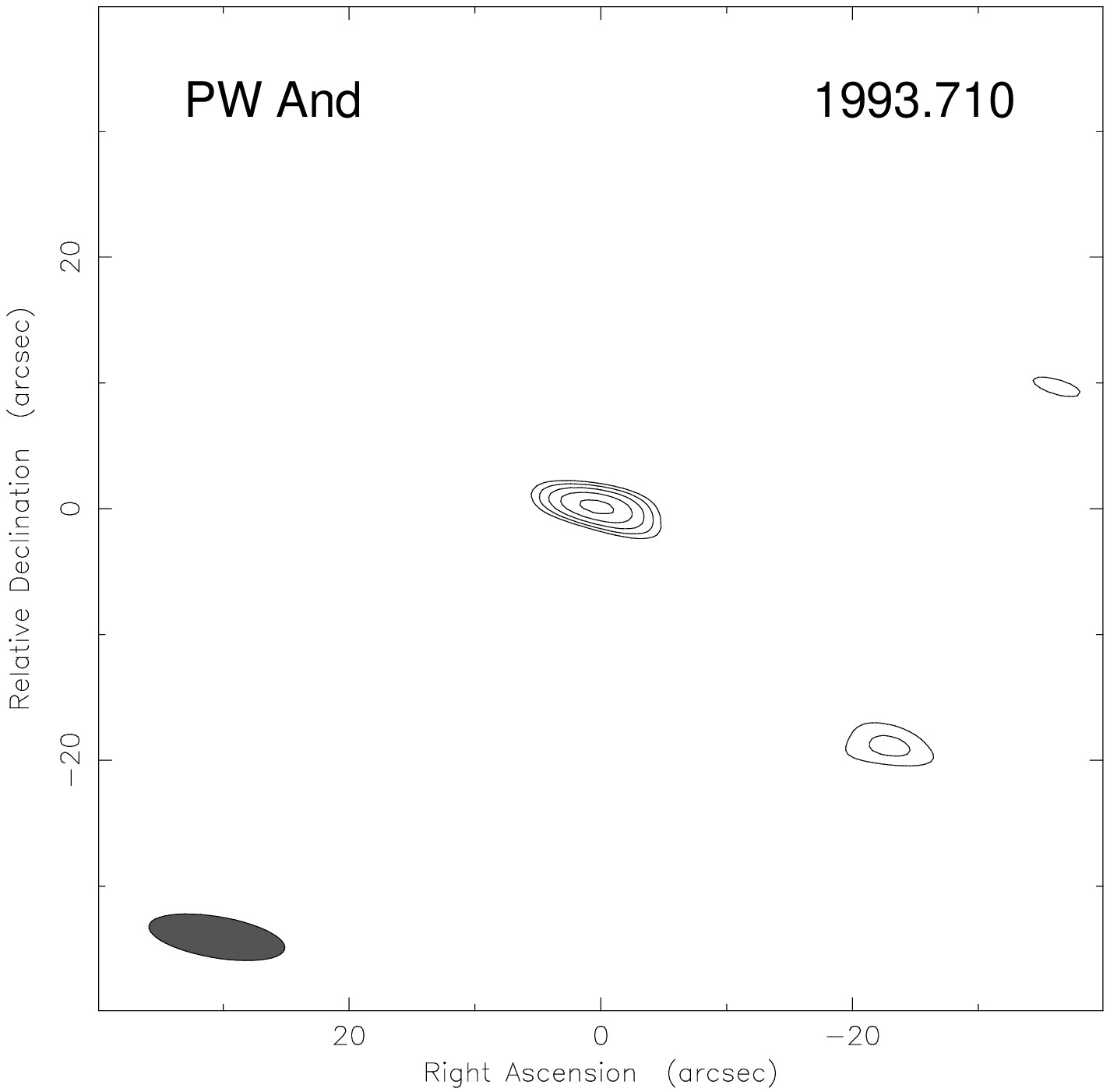}}
\caption[VLA map of the star PW\,And]{VLA 8.4\,GHz image of PW\,And taken on 1993.710. The lowest contour level corresponds to 3 times the statistical rms noise (0.02\,mJy\,beam$^{-1}$) with a scale factor between contiguous contours of $\sqrt{2}$. The peak flux density in the image is 0.29\,mJy/beam. The restoring beam (shown in the bottom-left corner of the map) is an elliptical Gaussian of $10.90\times3.28$\,arcsec (PA 80.$^{\circ}$5).}
\label{pwand_vla}
\end{center}
\end{figure}

\begin{figure}[h]
\begin{center}
\resizebox{0.8\hsize}{!}
{\includegraphics{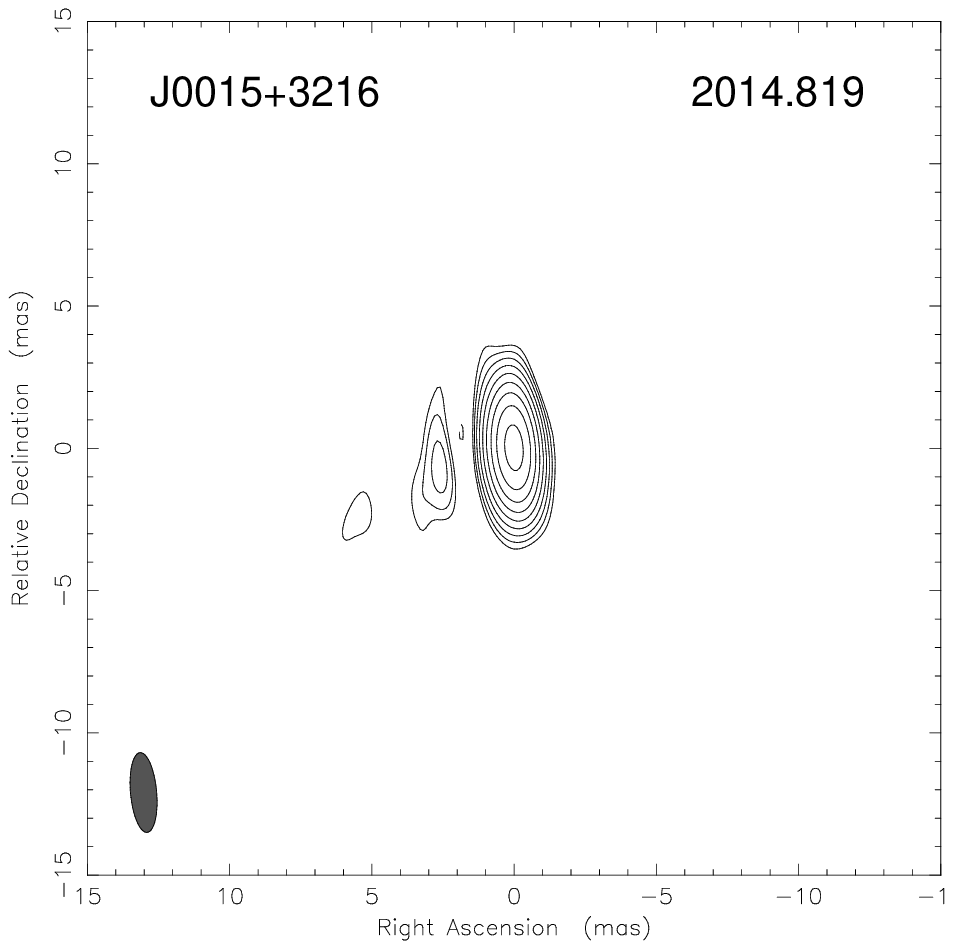}}
\caption[EVN maps of the calibrator J0015+3216]{EVN 5\,GHz image of J0015+3216 (calibrator of PW\,And) taken on 2014.819. The lowest contour level corresponds to 3 times the statistical rms noise (0.3\,mJy\,beam$^{-1}$) with a scale factor between contiguous contours of $\sqrt{3}$. The peak flux density in the image is 0.21\,mJy/beam.}
\label{j0015}
\end{center}
\end{figure}

\begin{figure}[h]
\begin{center}
\resizebox{0.8\hsize}{!}
{\includegraphics{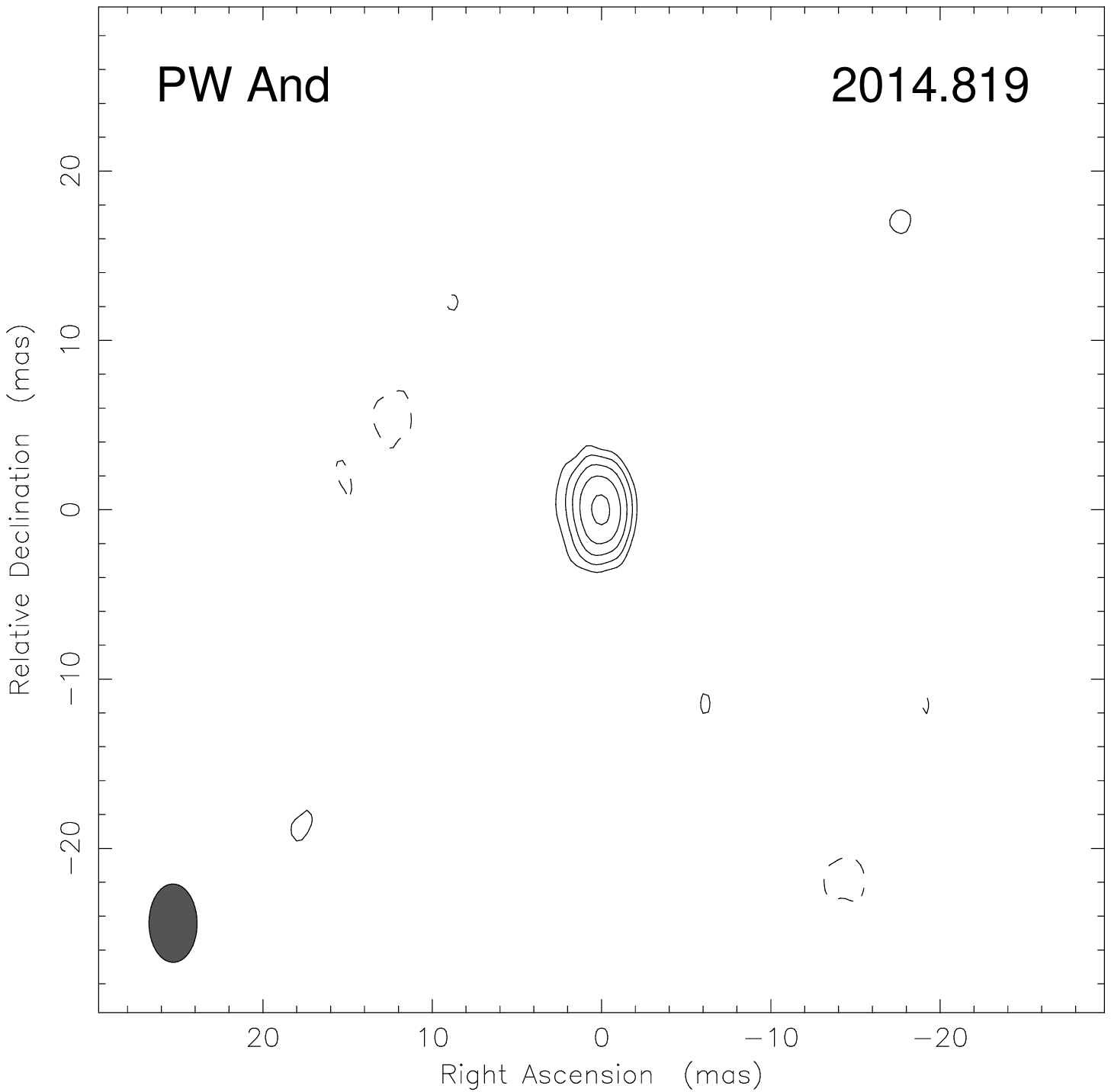}}
\caption[EVN map of the star PW\,And]{EVN 5\,GHz image of PW\,And taken on 2014.819. The lowest contour level corresponds to 3 times the statistical rms noise (0.01\,mJy\,beam$^{-1}$) with a scale factor between contiguous contours of $\sqrt{2}$. The peak flux density in the image is 0.16\,mJy\,beam$^{-1}$. The restoring beam (shown in the bottom-left corner of the map) is an elliptical Gaussian of $4.63\times2.84$\,mas (PA 0.$^{\circ}$40).}
\label{pwand}
\end{center}
\end{figure}

We used our AstraLux relative position of EK\,Dra to revisit the orbital motion between components A and B. This position is shown in Table \ref{poss_ek} along with already published relative positions of EK\,Dra\,A/B, which mostly results from speckle interferometry observations (K\"onig et al.~2005). We performed a weighted least-squares analysis similar to that presented for HD\,160934, simplified in this case to deal with only relative positions. Table \ref{tableorbitt_ek} shows the resulting orbital elements. The estimate of the combined mass of the system, using the \textit{Hipparcos} distance 33.94$\pm$0.72\,pc, is $m_{\mathrm{A}}$+$m_{\mathrm{B}}$ = 1.38$\pm$0.08\,M$_{\odot}$. Plots of the relative orbit can be seen in Fig.~\ref{o1_ek} and Fig.~\ref{o2_ek}. Both the orbital elements and mass estimates coincide with those reported by K\"onig et al.~(2005) within uncertainties. Although our new position extends twofold the time baseline of the orbital monitoring, the motion of B around the main star is very slow, indicating that the components are near the apoastron.

\begin{table*}
\caption{Compilation of astrometric measurements for the EK\,Dra system}             
\label{poss_ek}      
\begin{center}
%\resizebox{\hsize}{!}{       
\begin{tabular}{c c r r c }
\hline\hline
\noalign{\smallskip}
% & & Relative positions AB\,Dor\,Ba $-$ AB\,Dor\,Bb & & \\       
\multicolumn{5}{c}{Relative positions of EK\,Dra\,B $-$ EK\,Dra\,A} \\       
                      
Epoch & Instrument & $\Delta\alpha$\,(mas) & $\Delta\delta$\,(mas) & Reference \\ 
\hline                    
1991.2135 & 1D & +0.030$\pm$0.020 & $-$0.280$\pm$0.015 & (1) \\
1992.1232 & 1D & +0.045$\pm$0.022 & $-$0.310$\pm$0.015 & (1) \\
1993.7611 & MAGIC & +0.050$\pm$0.010 & $-$0.453$\pm$0.015 & (1) \\
1994.0712 & MAGIC & +0.051$\pm$0.004 & $-$0.483$\pm$0.010 & (1) \\
1994.9501 & MAGIC & +0.040$\pm$0.004 & $-$0.499$\pm$0.005 & (1) \\
1997.8926 & MAGIC & +0.059$\pm$0.008 & $-$0.644$\pm$0.012 & (1) \\
2001.1123 & OMEGA Cass & +0.085$\pm$0.007 & $-$0.674$\pm$0.005 & (1) \\
2001.8406 & OMEGA Cass & +0.074$\pm$0.013 & $-$0.705$\pm$0.012 & (1) \\
2002.8049 & OMEGA Cass & +0.102$\pm$0.012 & $-$0.718$\pm$0.009 & (1) \\
2013.4820 & AstraLux & +0.081$\pm$0.008 & $-$0.773$\pm$0.008 & (2) \\
\hline                  
\end{tabular}
%}
\end{center}
\footnotesize{\textbf{Notes.} (1) K\"onig et al.~(2005) using the 3.5\,m telescope on Calar Alto; (2) This work.}
\end{table*}

\begin{table}
\begin{center}
\caption[Estimates of orbital parameters of EK\,Dra]{Estimates of the orbital parameters of EK\,Dra}
\label{tableorbitt_ek}
\begin{tabular}{r l }
\hline\hline
Parameter & Value \\
\hline
$P$\,(yr): & 47.9 $\pm$ 0.9 \\
$a_{\mathrm{rel}}$\,($^{\prime\prime}$): & 0.434 $\pm$ 0.003 \\
$e$: & 0.812 $\pm$ 0.009 \\
$i\,(^\circ$): & 89 $\pm$ 1 \\
$\omega\,(^\circ$): & 180 $\pm$ 1  \\
$\Omega\,(^\circ$): & $-$186 $\pm$ 1\\
$T_{0}$: & 1986.2 $\pm$ 0.1 \\
\hline
\end{tabular}
\end{center}
%\footnotesize{\textbf{Notes.}The uncertainties shown are the standard errors scaled so that the $\chi^{2}$ per degree of freedom is unity.}
\end{table}

\subsection{PW\,Andromedae}
PW\,And (=HD\,1405) is a chomospherically very active star with a spectral type K2V, which displays a fast rotation (1.75 days) (Montes et al.~2001). Strassmeier et al.~(1988) included this object as a possible binary, however, radial velocity studies (Griffin 1992, L\'opez-Santiago et al.~2003) discard the presence of a close, interacting companion, showing that the chomospheric activity is due to PW\,And itself. Moreover, Evans et al.~(2012) explored the inner region of the star with speckle interferometry excluding the presence of companions at separations larger than 20\,mas.

Regarding the radio observations, a VLA image (Fig.~15) reveals a flux density of 0.34\,mJy. Our EVN image (Fig.~17), meanwhile, shows an unresolved source that should correspond to PW\,And with a flux density of 0.17\,mJy. Since our EVN observation does not show any companion to PW\,And, we can extend the absence of companions down to the resolution of our array, $\sim$5\,mas, at a flux density limit level of 0.01\,mJy. Given the apparent single character of the star, we did not propose further EVN observations of this source in the frame of this work (dedicated to monitor binary/multiple systems). Still, once confirmed the presence of compact emission, the determination of a precise, VLBI-based parallax value, which supersedes that of \textit{Hipparcos}, might be certainly of interest.

\subsection{LO\,Pegasus}
LO\,Peg (=BD+224402) is a young, active star with spectral type in the range K3V-K8V (Zuckerman \& Song 2004 Pandey et al.~2005) and a rotation period of 0.42 days (Barnes et al.~2005). The first study of LO\,Peg was carried out by Jeffries et al.~(1994), who concluded that there was no circumstellar matter around the star. Since then, Doppler images and studies of radial velocity have been carried out (Barnes et al.~2005, Piluso et al.~2008) and all of these studies consider LO\,Peg as a single star.

\begin{figure}[h]
\begin{center}
\resizebox{0.8\hsize}{!}
{\includegraphics{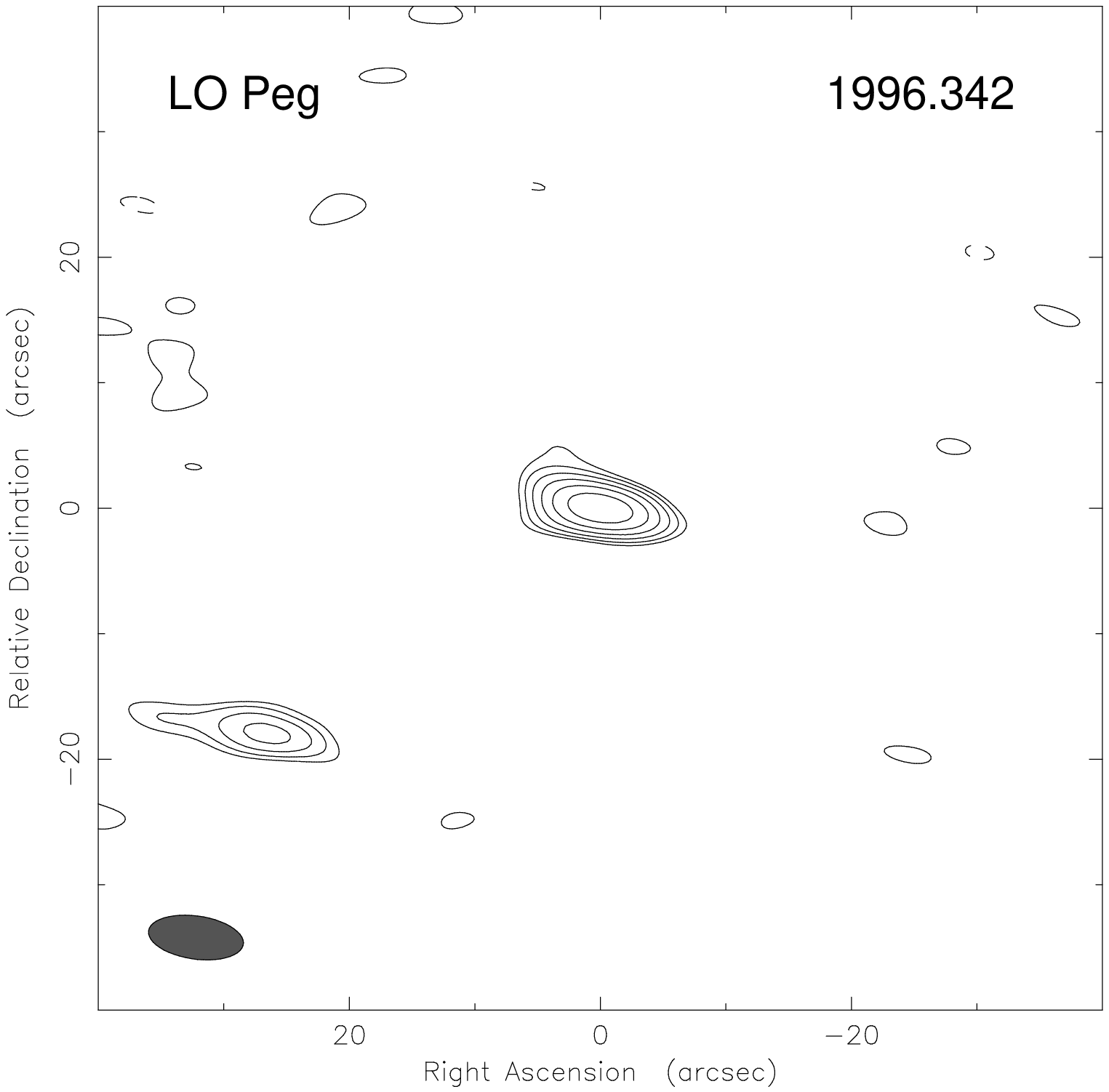}}
\caption[VLA map of the star LO\,Peg]{VLA 8.4\,GHz image of LO\,Peg taken on 1996.342. The lowest contour level corresponds to twice the statistical rms noise (0.02\,mJy\,beam$^{-1}$) with a scale factor between contiguous contours of $\sqrt{2}$. The peak flux density in the image is 0.42\,mJy/beam. The restoring beam (shown in the bottom-left corner of the map) is an elliptical Gaussian of $7.63\times3.44$\,arcsec (PA 82.$^{\circ}$0).}
\label{lopeg_vla}
\end{center}
\end{figure}

\begin{figure}[h]
\begin{center}
\resizebox{0.8\hsize}{!}
{\includegraphics{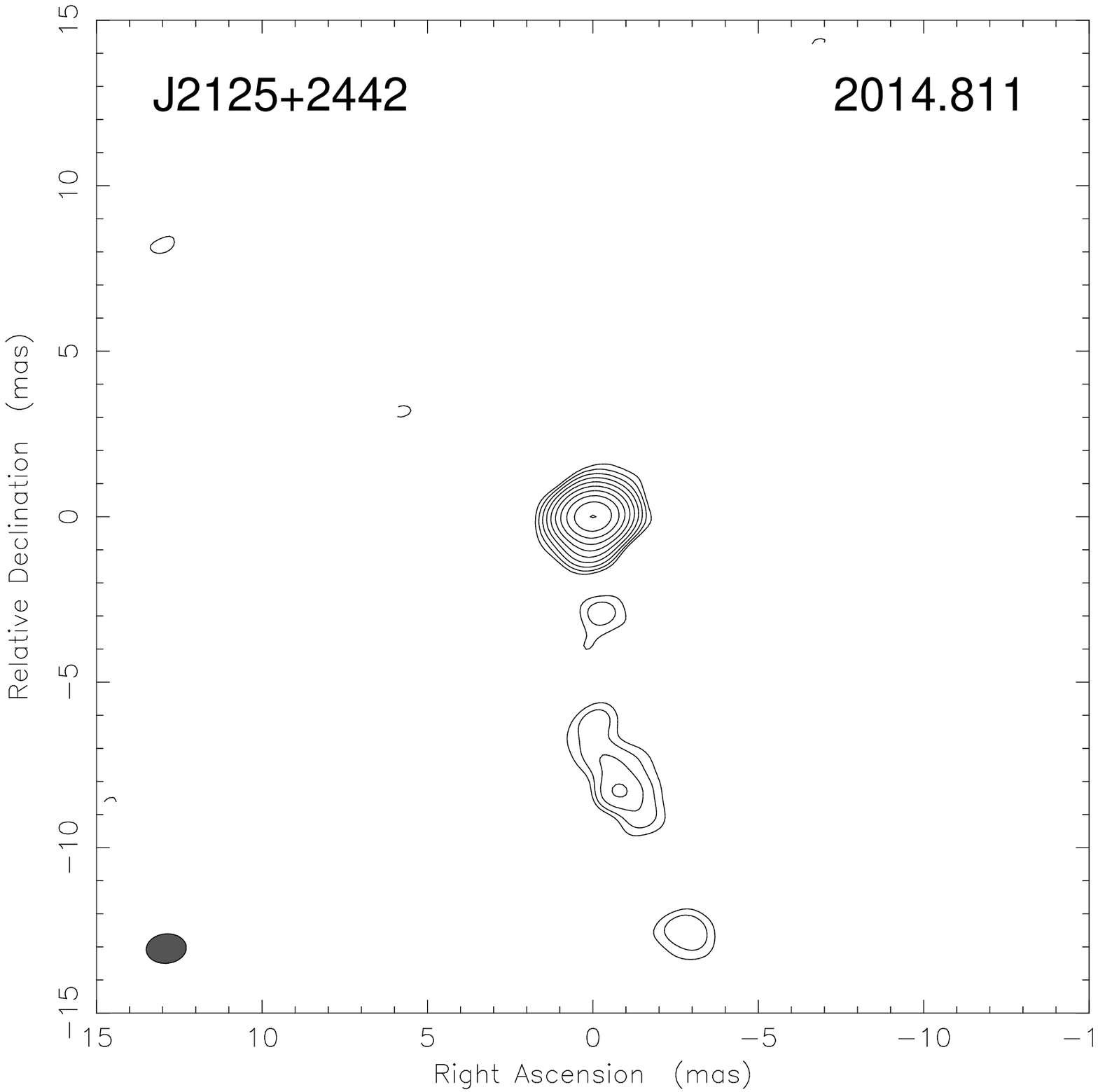}}
\caption[EVN maps of the calibrator J2125+2442]{EVN 5\,GHz image of J2125+2442 (calibrator of LO\,Peg) taken on 2014.811. The lowest contour level corresponds to 3 times the statistical rms noise (0.2\,mJy\,beam$^{-1}$) with a scale factor between contiguous contours of $\sqrt{3}$. The peak flux density in the image is 0.11\,mJy/beam.}
\label{j2125}
\end{center}
\end{figure}

We can confirm the radio emission of this source with a VLA image (Fig.~\ref{lopeg_vla}) that reveals a flux density of 0.45\,mJy. In the VLBI image, nevertheless, the star could not be detected. The upper bound to radio emission of LO\,Peg in this observation is 0.08\,mJy/beam. This is because the non-detection could reflect the high variability of these active stars.

\section{Discussion and conclusions}

\begin{table*}
\caption{Radio stars from the AB\,Doradus moving group}
\label{tablestars_mg}
\begin{center}
%\resizebox{\hsize}{!}{
\begin{tabular}{l c c c c c c} 
\hline \hline
Name & $T_{\mathrm{b}}$ & $v$\,sin\,$i$ &  P$_{\mathrm{rot}}$ & log$L_\mathrm{X}$ & log$L_\mathrm{R}$ & References\\ 
 & ($\times$10$^{6}$\,K) & (km/s) & (d) & (erg/s) & (erg/s/Hz) & \\
\hline
PW\,And & 0.81 & 24 & 1.75 & 30.5 & 14.5 & 1,12,13,14 \\
AB\,Dor\,Ba/Bb & 2.25 / 2.60 & 9 & 0.33 & $-$ & 14.6 & 2,3,4,5 \\
%AB\,Dor\,A/C & 2.25$\times$10$^{6}$ / $-$ & 44 & 0.514 & 30.2 - 32 & 15 - 16 & 2,3,4 \\ 
EK\,Dra\,A/B & 0.73 / $-$ & 16.5 & 2.78 & 29.92 & 14.6 & 1,6,7,14 \\
HD\,160934\,A/c & 0.72 / 1.4 & 17 & 1.8 & 29.39 & 14.4 & 2,8,9,10,14 \\
LO\,Peg & $-$ & 60 & 0.42 & 30.2 & 14.7 & 11,12,13,14 \\
\hline
\end{tabular}
%}
\end{center}
\footnotesize{\textbf{Notes.} (1) Montes el al.~(2001); (2) Zuckerman \& Song (2004); (3) Lim (1993); (4) Janson et al.~(2007); (5) Azulay et al.~(2015); (6) G\"udel et al.~(1995); (7) K\"onig et al.~(2005); (8) Hormuth et al.~(2007); (9) Evans et al.~(2012); (10) Azulay et al.~(2014); (11) Jeffries et al.~(1994); (12) Wichmann et al.~(2003); (13) VLA data archive; (14) This work.}
\end{table*}

The main properties of the AB\,Dor-MG stars studied in this paper are shown in Table \ref{tablestars_mg}, including our calculated values of the absolute radio luminosity (obtained from the VLA flux and the \textit{Hipparcos} distance) and the brightness temperature (obtained from the VLBI angular size when the source is detected). While AB\,Dor\,Ba/Bb and HD\,160934\,A/c resulted to be intense radio emitters that were detected at all observing epochs (and so it is PW\,And, also detected in our unique VLBI epoch of observation), we found that EK\,Dra (in 2 out 3 epochs) and LO\,Peg did not display detectable levels of radio emission. These non-detections might just reflect the variability of the radio emission, since, as seen in Table \ref{tablestars_mg}, neither the distance, rotation period, nor X-ray luminosity are significantly different in these two stars with respect to the other radio emitter systems. Therefore, further monitoring of these non-detected stars would not be as efficient in terms of kinematical studies as in the cases presented for AB\,Dor\,Ba/Bb (Azulay et al. 2015) and HD\,160934 (this paper). The same conclusion applies to PW\,And, given its apparent non-binarity (even taking into account a clear VLBI detection). 

On the other hand, from the brightness temperatures shown in Table \ref{tablestars_mg}, we can conclude that the radio emission has a non-thermal origin. This fact, along with the rapid rotation values and saturated levels of X-ray luminosity, $L_\mathrm{X}$, also listed in the table, favors the existence of an intense magnetic activity of the stellar corona that is responsible in terms of the presence of this radio emission. Therefore, the radio emission is apparently generated by gyrosynchrotron emission from non-thermal accelerated electrons (Lim et al. 1994; G\"udel et al.~1995).

In this paper we have shown the results of a VLBI program dedicated to monitoring the absolute reflex motion of HD\,160934, a member of the AB\,Dor-MG. The unexpected detection of compact radio emission of the low-mass companion c (Azulay et al. 2014) allowed us to sample not only the absolute orbit of component A (with respect to the external quasar J1746+6226), but also the relative orbit, both of which are necessary to determine model-independent dynamical masses of the components of this system. The proximity of the two stars near periastron ($\sim$20$-$40\,mas in the last four years) has prevented an appropriate sampling of the relative orbit until the recent use of more precise interferometric techniques: aperture-masking (Evans et al.~2012) and VLBI (this work). The results of our orbital analysis yields values of 0.70$\pm$0.07\,M$_{\odot}$ and 0.45$\pm$0.04\,M$_{\odot}$ for components A and c, respectively, which are larger than the theoretical values predicted by PMS evolutionary tracks. The amount of this disagreement is $\sim$10\% for component A and 10$-$40\% for component c, contributing to the increasing observational evidence that PMS models underpredict the masses of systems with masses below 1.2\,M$_\odot$. We have found that the inclusion of the effect of the stellar magnetic field in the theoretical models tends to reduce such a discrepancy. Remarkably, our study allowed us to obtain a revised and more precise value of the parallax (31.4$\pm$0.5\,mas), thus solving a long-standing discussion about the distance to this system.

With respect to the other stars included in our study, EK\,Dra and PW\,And showed compact radio emission at milliarcsecond scales, meanwhile LO\,Peg, appeared to be ``off'' at the time of observations. Complementary, companion infrared observations of EK\,Dra permitted us a revision of the orbital parameters of this system.

\begin{acknowledgements}
R.A., J.C.G., J.M.M., \& E.R. were partially supported by the Spanish MINECO projects AYA2012-38491-C02-01 and AYA2015-63939-C2-2-P and by the Generalitat Valenciana project PROMETEO/2009/104 and PROMETEOII/2014/057. R.A.\ acknowledges the Max-Planck-Institute f\"ur Radioastronomie for its hospitality. E.T. was supported by the ``PRA 2016 Universit\`a di Pisa''. E.T. also acknowledges the INFN iniziativa specifica TAsP.
\end{acknowledgements}

\end{document}